\documentclass[10pt,final,journal,twocolumn]{IEEEtran}
\ifCLASSINFOpdf

\else
\usepackage{algorithm}
\usepackage{booktabs}

\fi
\usepackage{cite}
\usepackage[cmex10]{amsmath}
\usepackage{amsthm}
\usepackage{stfloats}
\usepackage{multicol,multienum}
\usepackage{multirow}
\usepackage{mdframed}
\usepackage{amssymb}
\usepackage{url}
\usepackage{amsmath}
\usepackage{xcolor}
\usepackage[dvips]{graphicx}
\usepackage{subfigure}
\usepackage{caption}
\usepackage{amsmath}
\usepackage{amsfonts,amssymb}
\usepackage{bbm}
\usepackage[T1]{fontenc}
\usepackage{algorithm}
\usepackage{algpseudocode}
\usepackage{ulem}
\usepackage{balance}

\usepackage{caption}
\usepackage[linesnumbered, ruled]{}
\makeatletter
\renewcommand{\maketag@@@}[1]{\hbox{\m@th\normalsize\normalfont#1}}%
\makeatother

\begin{document}
\hyphenation{op-tical net-works semi-conduc-tor}
\title{Subverting Flexible Multiuser Communications via Movable Antenna-Enabled Jammer}
\author{Guojie Hu, Qingqing Wu,~\textit{Senior Member}, \textit{IEEE}, Lipeng Zhu,~\textit{Member}, \textit{IEEE}, Kui Xu,~\textit{Member}, \textit{IEEE}, Guoxin Li, Jiangbo Si,~\textit{Senior Member}, \textit{IEEE}, Jian Ouyang,~\textit{Member}, \textit{IEEE}, and Tong-Xing Zheng,~\textit{Member}, \textit{IEEE}\vspace{-1.3em}
\thanks{
%
Guojie Hu is with the Department of Electronic Engineering, Shanghai Jiao Tong University, Shanghai 200240, China, and also with the College of Communication Engineering, Rocket Force University of Engineering, Xi'an
710025, China (lgdxhgj@sina.com). Qingqing Wu is with the Department of Electronic Engineering, Shanghai Jiao Tong University, Shanghai 200240, China. Lipeng Zhu is with the Department of Electrical and Computer Engineering, National University of Singapore, Singapore 117583. Kui Xu and Guoxin Li are with the College of Communications Engineering, Army Engineering University of PLA, Nanjing 210007, China. Jiangbo Si is with the Integrated Service Networks Lab of Xidian University, Xi'an 710100, China. Jian Ouyang is with the Institute of Signal Processing and Transmission, Nanjing University of Posts and Telecommunications, Nanjing 210003, China. Tong-Xing Zheng is with the School of Information and Communications Engineering, Xi'an Jiaotong University, Xi'an 710049.
}
}
\IEEEpeerreviewmaketitle
\maketitle
\begin{abstract}
Movable antenna (MA) is an emerging technology which can reconfigure wireless channels via adaptive antenna position adjustments at transceivers, thereby bringing additional spatial degrees of freedom for improving system performance. In this paper, from a security perspective, we exploit the MA-enabled legitimate jammer (MAJ) to subvert suspicious multiuser downlink communications consisting of one suspicious transmitter (ST) and multiple suspicious receivers (SRs). Specifically, our objective is to minimize the benefit (the sum rate of all SRs or the minimum rate among all SRs) of such suspicious communications, by jointly optimizing antenna positions and the jamming beamforming at the MAJ. However, the key challenge lies in that given the MAJ's actions, the ST can reactively adjust its power allocations to instead maximize its benefit for mitigating the unfavorable interference. Such flexible behavior of the ST confuses the optimization design of the MAJ to a certain extent. Facing this difficulty, corresponding to the above two different benefits: i) we respectively determine the optimal behavior of the ST given the MAJ's actions; ii) armed with these, we arrive at two simplified problems and then develop effective alternating optimization based algorithms to iteratively solve them. In addition to these, we also focus on the special case of two SRs, and reveal insightful conclusions about the deployment rule of antenna positions at the MAJ. Furthermore, we analyze the ideal antenna deployment scheme at the MAJ for achieving the globally performance lower bound. Numerical results demonstrate the effectiveness of our proposed schemes compared to conventional fixed-position antenna (FPA) and other competitive benchmarks.
\end{abstract}

\IEEEpeerreviewmaketitle
\vspace{-10pt}
\section{Introduction}
The rapid evolution of wireless communications has promoted prosperous developments of various industries, such as agriculture, education, business, etc. However, the great portability of wireless communications has also given risen to potential risks where terrorists and commercial criminals can exploit it at anytime and anywhere to commit crimes, endanger public security, etc. \cite{Ke_wen}. Therefore, from the perspective of the legitimate parties, one urgent task is how to effectively suppress such kind of suspicious communications \cite{jamm_game}.

Wireless jamming, as an attractive approach for deteriorating the signal reception quality at suspicious receivers (SRs), has received widespread attention for handling the above issue. For instance, via careful jamming power allocations in the training and data transmission phases, \cite{uplink_MIMO}, \cite{uplink_MIMO_CRN} and \cite{cell_free} aimed to maximally reduce the sum spectral efficiency of uplink massive multiple-input-multiple-output (MIMO) systems, joint MIMO and cognitive radio networks (CRNs), and user-centric cell-free networks, respectively. The work \cite{worst_case} designed the worst-case jamming scheme on a point-to-point MIMO system consisting of one suspicious transmitter (ST) and one SR, under the background that the jammer knows the global channel state information (CSI). Compared to \cite{worst_case}, \cite{Nan_Zhao} further considered that the ST and the SR can adaptively adjust their strategies to moderate the unfavorable jamming. Under this setup, \cite{Nan_Zhao} investigated the optimal jamming design under different cases about the availability of CSI at the jammer. Subsequently, \cite{jamming_relay} studied the optimal jamming on the relay-assisted multiuser communications and transformed the jamming/anti-jamming game as the min-max problem. In addition, \cite{user_scheduling} considered that multiple jammers cooperatively disturbed the uplink channel training results of one suspicious MIMO multiuser system, in order to minimize the number of downlink scheduled users.

 \begin{figure*}[!t]
\centering
\includegraphics[width=13cm]{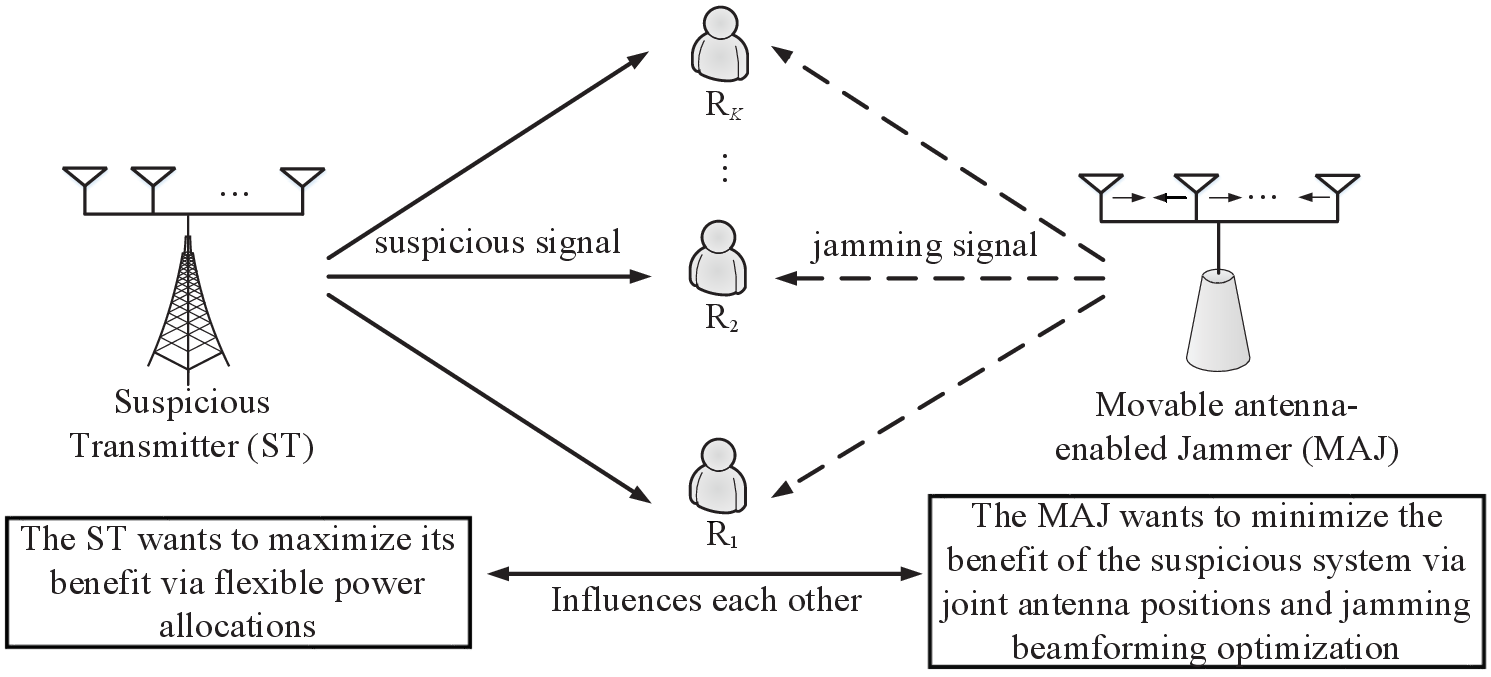}
\captionsetup{font=small}
\caption{Illustration of the considered system model.} \label{fig:Fig1}
\end{figure*}

Current jamming almost relies on the fixed-position antenna (FPA) array, where the distance between arbitrary adjacent antennas at the jammer is fixed as the half wavelength. Under such setting, jamming channels cannot be proactively reconfigured. Therefore, in the situation where jamming channel conditions are unfavorable, the jamming beamforming may not be able to fully unleash its potential. To overcome this deficiency, recent works proposed a novel movable antenna (MA) technology \cite{ZLP_survey}, where the implementation methods for MAs can be broadly categorized into two approaches: mechanical control, which physically repositions antenna elements or arrays using actuators or structures, and electronic control, which equivalently adjusts antenna positions or radiation patterns by exciting different electromagnetic modes or reconfiguring dense arrays \cite{Ningboyu_WCM}. Compared to the FPA technology, the MA technology reaps an additional spatial degree of freedom (DoF), i.e., it can flexibly reconfigure wireless channels via antenna position movements to create favorable prerequisites for flexible beamforming \cite{Z_WCM}.

Attracted by its potential advantages, early works have considered the application of the MA technology in various scenarios. For instance, \cite{ZLP_MODEL} proposed the field-response channel model tailored to MA systems and analyzed the maximum channel gain achieved by a single MA. Subsequently, \cite{MWYY_TWC,Gao_Xiqi_CL,Yongpeng_WU_Glob} investigated the MIMO system where the transmitter and the receiver all equip with multiple MAs. By jointly optimizing positions of MAs at transceivers and the covariance of transmit signals, the MA technology shows its significant gains in improving the capacity compared to the FPA technology. In \cite{XZY_TWC,Guojie_Fluid,Songjie_WCL,MA_added}, multiuser communications empowered by the MA technology were considered, where antenna positions at users or the base station and the transmitting/receiving beamforming are jointly optimized to improve the sum rate or reduce the total transmit power. Unlike most works on continuously searching for optimizing MA positions, \cite{MWD1} and \cite{MWD2} proposed to sample the transmit region into discrete points and then developed the graph theory based algorithm to determine the optimal and discrete antenna positions deployments in polynomial time. To further improve the spatial DoF, \cite{6DMA1} and \cite{6DMA2} proposed a six-dimensional MA (6DMA) architecture, where in addition to 3D positions of all antennas, 3D rotations of antenna surfaces can be designed to achieve the maximum flexibility. Besides, there are other works that envisaged numerous MAs-enabled setups, such as near-filed communications \cite{Near_field}, downlink multicasting \cite{multicast}, unmanned aerial vehicle communications \cite{UAV_MA}, integrated sensing and communication (ISAC) \cite{ISAC1,ISAC2}, mobile edge computing \cite{MEC_MA}, security and covert communications \cite{Guojie_SPL,covert_MA}, etc.

To date, while existing works \cite{Guojie_SPL,covert_MA} have applied the MA technology to enhance secure wireless communications, conversely, the potential for exploiting this technology to effectively suppress the growing threat of suspicious communications remains largely unexplored and warrants significant attentions. Crucially, the MA technology offers unique advantages for jamming applications: by dynamically adjusting the positions of its antennas, the jammer can continuously optimize the spatial channel characteristics. This enables the precise manipulation of jamming channels towards the suspicious receivers, potentially achieving significantly stronger and more focused jamming effects compared to FPAs. Motivated by this observation, we consider in this paper a new setup where the MA-enabled legitimate jammer (MAJ) aims to minimize the benefit of the suspicious multiuser downlink communications consisting of one ST and multiple SRs, by jointly optimizing its jamming beamforming and antenna positions. Here, we consider two typical benefits for the suspicious communications, i.e., the sum rate of all SRs and the minimum rate among all SRs. Note that in the considered model, the key challenge lies in that given the actions of the MAJ, the ST could adaptively optimize its power allocations to moderate the unfavorable interference for maximizing the benefit of itself. Obviously, such flexible reaction of the ST confuses the MAJ about the exact interference effect it caused and also increases the optimization difficulties from the perspective of the MAJ. Facing this challenge, we provide corresponding methods for simplifying and solving the problems, which are summarized as follows.
\begin{itemize}
\item[$\bullet$] In Scenario I where the suspicious system cares about maximizing the sum rate, the optimal water-filling power allocations of the ST given the actions of the MAJ are first derived, based on which the problem is simplified into one only related to the jamming beamforming, the antenna positions at the MAJ, and the water-filling level. However, the constraints still include the complex $\max ( \cdot )$ function, the low bound of which is hard to be derived and thus will hinder the subsequent optimizations. To tackle this, the log-sum-exp function is employed to tightly approximate the $\max ( \cdot )$ function, based on which the alternating optimization (AO) and the successive convex approximation (SCA) are successfully designed to iteratively update the above variables until convergence.

 \item[$\bullet$] In Scenario II where the suspicious system cares about maximizing the minimum rate, the optimal power allocations of the ST given the actions of the MAJ are derived in closed forms, substituting which into the objective and via certain transformations, the problem can be equivalent to maximizing the principal eigenvalue of a hermite matrix which is only related to antenna positions at the MAJ. However, since the expression of the principal eigenvalue is much complex, we further provide an equivalent alternative with much clear structures to replace it. Based on these, we then develop the easy-to-implement AO and SCA based algorithms to iteratively solve the simplified problem.

 \item[$\bullet$] To gain insightful conclusions, we further focus on the special case of two SRs and strictly prove that antenna positions and the jamming beamforming at the MAJ can be optimized separately in this case. Specifically, in Scenarios I and II, antenna positions should be first optimized based on the rule of maximizing the correlation of two jamming channels, the process of which has nothing to do with the jamming beamforming. Afterwards, in Scenario I, the optimal jamming beamforming can be obtained via the simple search; and in Scenario II, it can be derived with the closed-form solution.

 \item[$\bullet$] We also analyze the necessary conditions for achieving the performance lower bound in both Scenarios, i.e., the correlation between any two jamming channels reaches the global maximum. Based on this insight, we further derive the ideal antenna position deployments at the MAJ for satisfying the above conditions.
\end{itemize}

The rest of this paper is organized as follows. Section II introduces the system model and then presents the problem formulation. Section III and IV provide the detailed optimization processes for solving the problems corresponding to Scenarios I and II, respectively. Section V focuses on the special case of two SRs and then derives insightful conclusions about the deployment rule of antenna positions at the MAJ. Section VI presents the ideal antenna position deployments at the MAJ for achieving the performance bounds. Simulation results are provided in Section VII and finally the conclusions are given in Section VIII.

{\textit{{Notations:}} For a complex number $a$, ${\mathop{\rm Re}\nolimits} \left[ a \right]$ and $\left| a \right|$ denote its real part and amplitude, respectively. For a complex vector ${\bf{h}}$, ${{\bf{h}}^T}$, ${{\bf{h}}^*}$, ${{\bf{h}}^H}$ and $\left\| {\bf{h}} \right\|$ denote its transpose, conjugate, conjugate transpose and Frobenius norm, respectively. ${{\bf{I}}_K}$ denotes the identity matrix of order $K$. For a matrix ${\bf{A}}$, ${\rm{Tr}}({\bf{A}})$ and ${{\bf{A}}^{ - 1}}$ denote the its trace and inverse, respectively. ${\cal CN}(0,\Delta )$ denotes the circularly symmetric complex Gaussian (CSCG) distribution with mean zero and covariance $\Delta$.}

 %


 \newcounter{mytempeqncnt}
\section{System Model and Problem Formulation}
\subsection{System Model}
As shown in Fig. 1, we consider a challenging scenario consisting of two opposing parties. The suspicious party includes $K$ single-antenna SRs (denoted as $\left\{ {{{\rm{R}}_i}} \right\}_{i = 1}^K$), and one ST equipped with a linear FPA array, which is composed of $M$ antennas. The legitimate party includes one MAJ equipped with a linear MA array, which is composed of $N$ antennas. Specifically,
\begin{itemize}
\item[$\bullet$] For the suspicious party, the ST adopts the zero-forcing (ZF)-based transmit beamforming and concurrently exploits the adaptive power allocations to maximize the benefit of itself. To ensure comprehensiveness, we will respectively consider two typical benefits in this paper, which are: i) the sum rate of $K$ users; ii) the minimum rate among $K$ users.

 \item[$\bullet$] While for the legitimate party, the MAJ aims to minimize the benefit of the suspicious communications by emitting jamming signals to interfere with $\left\{ {{{\rm{R}}_i}} \right\}_{i = 1}^K$. To achieve this purpose more effectively, the MAJ will
     jointly optimize antenna positions and the jamming beamforming at itself. The details are shown later.
\end{itemize}

\setcounter{equation}{\value{mytempeqncnt}}
  \begin{figure*}[b!]
  \hrulefill
\setcounter{mytempeqncnt}{\value{equation}}
\setcounter{equation}{13}
\begin{equation}
\begin{split}{}
{R_{{\rm{sum}}}}(\eta ,{{\bf{w}}_J}) = \frac{1}{{\ln 2}}\sum\nolimits_{i = 1}^K {\max \left( {\underbrace {\ln \left( {{{\left| {{\bf{w}}_{i,{\rm{ZF}}}^H{{\bf{h}}_i}} \right|}^2}} \right)}_{{{\rm{C}}_{1,i}}} + \underbrace {\ln \eta }_{{\Theta _1}(\eta )} - \ln \left( {{\sigma ^2} + {Q_J}\underbrace {{{\left| {{\bf{w}}_J^H{{\bf{g}}_i}({\bf{x}})} \right|}^2}}_{{\Theta _{2,i}}({{\bf{w}}_J})}} \right),0} \right)} .
\end{split}
\end{equation}
\setcounter{equation}{\value{mytempeqncnt}}
\vspace{-12pt}
\end{figure*}

To simplify analysis and without affecting the obtained conclusions, we consider the line-of-sight (LoS) channel model for all involved links. Specifically, the channel vector from the ST to each ${\rm{R}}_i$ ($i = 1,...,K$) can be expressed as
\setlength\abovedisplayskip{1.65pt}
\setlength\belowdisplayskip{1.65pt}
\begin{equation}
\begin{split}{}
{{\bf{h}}_i} = \frac{1}{{\sqrt {d_{{S_i}}^\tau } }}{\left[ {1,{e^{j\frac{{2\pi }}{\lambda }d\sin {\theta _i}}},...,{e^{j\frac{{2\pi }}{\lambda }(M - 1)d\sin {\theta _i}}}} \right]^T},
\end{split}
\end{equation}
where ${d_{{S_i}}}$ is the distance between the ST and ${\rm{R}}_i$, $\tau $ is the path loss exponent, $\lambda $ is the signal wavelength, $d \buildrel \Delta \over = \lambda /2$ is the fixed spacing of arbitrary two adjacent antennas for the FPA array, and ${{\theta _i}}$ is the angle of arrival (AoA) to the ST at ${\rm{R}}_i$.

At the same time, unlike ${{\bf{h}}_i}$, the channel vector from the MAJ to each ${\rm{R}}_i$ ($i = 1,...,K$) will be related to the adjustable antenna positions at the MAJ, which can be expressed as
\begin{equation}
\begin{split}{}
{{\bf{g}}_i}({\bf{x}}) = \frac{1}{{\sqrt {d_{{J_i}}^\tau } }}{\left[ {{e^{j\frac{{2\pi }}{\lambda }{x_1}\sin {\varphi _i}}},...,{e^{j\frac{{2\pi }}{\lambda }{x_N}\sin {\varphi _i}}}} \right]^T},
\end{split}
\end{equation}
where ${\bf{x}} \buildrel \Delta \over = \left[ {{x_1},...,{x_N}} \right]$ and $x_i$ is the one-dimensional position of the $i$-th antenna relative to the reference point zero, ${d_{{J_i}}}$ is the distance between the MAJ and ${\rm{R}}_i$, and ${{\varphi _i}}$ is the AoA to the MAJ at ${\rm{R}}_i$.

Based on the above analysis, the signal received at ${\rm{R}}_i$ is given by
\begin{equation}
\begin{split}{}
{y_i} = \sqrt {{P_i}} {\bf{w}}_{i,{\rm{ZF}}}^H{{\bf{h}}_i}{s_i} + \sqrt {{Q_J}} {\bf{w}}_J^H{{\bf{g}}_i}({\bf{x}}){s_J} + {n_i},
\end{split}
\end{equation}
where $P_i$ and ${{\bf{w}}_{i,{\rm{ZF}}}} \in {{\mathbb{C}}^{M \times 1}}$ are the power and the ZF-based beamforming of the ST for transmitting the suspicious signal ${s_i} \sim {\cal C}{\cal N}(0,1)$ to ${\rm{R}}_i$, respectively, and ${{\bf{w}}_{i,{\rm{ZF}}}}$ with $\left\| {{{\bf{w}}_{i,{\rm{ZF}}}}} \right\| = 1$ has the closed-form expression as ${{\bf{w}}_{i,{\rm{ZF}}}} = \frac{{\left( {{{\bf{I}}_M} - {{\bf{A}}_i}{{({\bf{A}}_i^H{{\bf{A}}_i})}^{ - 1}}} \right){{\bf{h}}_i}}}{{\left\| {\left( {{{\bf{I}}_M} - {{\bf{A}}_i}{{({\bf{A}}_i^H{{\bf{A}}_i})}^{ - 1}}} \right){{\bf{h}}_i}} \right\|}}$ \cite{Guo_twotimescale}, where ${{\bf{A}}_i} = \left[ {{{\bf{h}}_1},...,{{\bf{h}}_{i - 1}},{{\bf{h}}_{i + 1}},...,{{\bf{h}}_K}} \right] \in {{\mathbb{C}}^{M \times (K - 1)}}$. Here, since ${\left| {{\bf{w}}_{j,{\rm{ZF}}}^H{{\bf{h}}_i}} \right|^2} = 0$, $\forall j \in \left\{ {1,...,K} \right\}$ and $j \ne i$, the multiuser interference is not reflected in (3). In addition, $Q_J$ is the jamming power budget of the MAJ, ${{\bf{w}}_J} \in {{\mathbb{C}}^{N \times 1}}$ is the jamming beamforming with $\left\| {{{\bf{w}}_J}} \right\| = 1$, ${s_J} \sim {\cal C}{\cal N}(0,1)$ is the jamming signal and ${n_i} \sim {\cal C}{\cal N}(0,{\sigma ^2})$ is the additive white Gaussian noise (AWGN) at ${\rm{R}}_i$.

\textbf{Remark}: Here, we consider the ZF-based transmit beamforming for the ST primarily for two reasons: i) it offers the low implementation complexity, especially for large $M$; and ii) with the fixed ZF-based beamforming, although the ST can optimize its power allocations to mitigate adverse interference from the MAJ, this power-domain adaptation can be efficiently inferred by the MAJ. This predictability facilitates the MAJ's optimization design to some extent. Conversely, if the ST further adopts the transmit beamforming with other forms such as weighted minimum mean square error (WMMSE), there may not exist the optimal and closed-form power allocations of the ST. Therefore, it becomes significantly difficult for the MAJ to determine the ST's exact response to its own actions. This would pose substantial challenges in optimization designs from the perspective of the MAJ.

 Accordingly, based on (3), the signal-to-interference-plus-noise ratio (SINR) of ${\rm{R}}_i$, which is the function of $P_i$, ${{\bf{w}}_J}$ and ${\bf{x}}$, can be expressed as
\begin{equation}
\begin{split}{}
{\gamma _i}({P_i},{{\bf{w}}_J},{\bf{x}}) = \frac{{{P_i}{{\left| {{\bf{w}}_{i,{\rm{ZF}}}^H{{\bf{h}}_i}} \right|}^2}}}{{{Q_J}{{\left| {{\bf{w}}_J^H{{\bf{g}}_i}({\bf{x}})} \right|}^2} + {\sigma ^2}}}.
\end{split}
\end{equation}

Hence, the sum rate or the minimum rate of the suspicious communications can be respectively given by
\begin{equation}
\begin{split}{}
{R_{{\rm{sum}}}}(\left\{ {{P_i}} \right\},{{\bf{w}}_J},{\bf{x}}) =& \sum\limits_{i = 1}^K {{{\log }_2}\left( {1 + {\gamma _i}({P_i},{{\bf{w}}_J},{\bf{x}})} \right)}, \\
{R_{{\rm{min}}}}(\left\{ {{P_i}} \right\},{{\bf{w}}_J},{\bf{x}}) =& \mathop {\min }\limits_i \left\{ {{{\log }_2}\left( {1 + {\gamma _i}({P_i},{{\bf{w}}_J},{\bf{x}})} \right)} \right\}.
\end{split}
\end{equation}

\subsection{Problem Formulation}
Standing on the side of the MAJ, it aims to minimize the benefit of the suspicious communications by jointly optimizing ${{\bf{w}}_J}$ and ${\bf{x}}$. That is to say, if the suspicious system cares about the sum rate maximization, the MAJ will accordingly minimize such sum rate; otherwise, if the suspicious system cares about the fairness, the MAJ will instead try to reduce the minimum rate. However, in both scenarios, the key difficulty lies in that given ${{\bf{w}}_J}$ and ${\bf{x}}$, the ST can adaptively adjust its power allocations $\left\{ {{P_i}} \right\}_{i = 1}^K$ to moderate the unfavorable interference for maximizing the benefit of itself. Here, $\sum\nolimits_{i = 1}^K {{P_i}}  \le {P_{{\rm{sum}}}}$ with ${P_{{\rm{sum}}}}$ denoting the power budget of the ST. Based on this fact, the optimization problems corresponding to the above two scenarios should be formulated as
 \begin{align}
&({\rm{P1}}):{\rm{  }}\mathop {\min }\limits_{{{{\bf{w}}_J},{\bf{x}}}} \ {\rm{Benefit}}\left( {\left\{ {P_i^*} \right\},{{\bf{w}}_J},{\bf{x}}} \right) \tag{${\rm{6a}}$}\\
{\rm{              }}&{\rm{s.t.}} \ \left\| {{{\bf{w}}_J}} \right\| \le 1,\tag{${\rm{6b}}$}\\
 &\ \ \quad {x_n} - {x_{n - 1}} \ge {d_{\min }},n = 2,3,...,N,\tag{${\rm{6c}}$}\\
 &\ \ \ \ \ \left\{ {{x_n}} \right\}_{n = 1}^N \in [0,L],\tag{${\rm{6d}}$}\\
 &\ \ \ \ \left\{ {P_i^*} \right\}_{i = 1}^K = \arg \mathop {\max }\limits_{\sum\limits_{i = 1}^K {{P_i}}  \le {P_{{\rm{sum}}}}} {\rm{Benefit}}\left( {\left\{ {P_i} \right\},{{\bf{w}}_J},{\bf{x}}} \right) ,\tag{${\rm{6e}}$}
 \end{align}
 where ${\rm{Benefit}}\left( {\left\{ {P_i} \right\},{{\bf{w}}_J},{\bf{x}}} \right) = {R_{{\rm{sum}}}}(\left\{ {P_i} \right\},{{\bf{w}}_J},{\bf{x}})$ if the suspicious system cares about the sum rate maximization, and ${\rm{Benefit}}\left( {\left\{ {P_i} \right\},{{\bf{w}}_J},{\bf{x}}} \right) = {R_{{\rm{min}}}}(\left\{ {P_i} \right\},{{\bf{w}}_J},{\bf{x}})$ if the suspicious system cares about the minimum rate maximization. In addition, ${d_{\min }}$ in (6c) is the minimum spacing between any two adjacent MAs for avoiding the coupling effect, $L$ in (6d) is the total span for the movement region of MAs, and the constraint in (6e) indicates the flexible power allocations of the ST given the MAJ's actions.

Generally, (P1) is hardly solved since: i) the independent variables ${{{\bf{w}}_J}}$ and ${\bf{x}}$ are highly coupled with each other in the objective; ii) the dependent variables $\left\{ {P_i^*} \right\}_{i = 1}^K$ adaptively change w.r.t. ${{{\bf{w}}_J}}$ and ${\bf{x}}$ shown in (6e). Such flexible behavior of the ST will in turn confuse the MAJ about the exact interference effect it caused to the suspicious system. In the next section, focusing on two different expressions of ${\rm{Benefit}}\left( {\left\{ P_i \right\},{{\bf{w}}_J},{\bf{x}}} \right)$, we will respectively develop the AO based algorithms to iteratively solve (P1).

\subsection{Extension to the Rician Channel}
If the Rician channel is considered, the channel vector from the ST to each ${\rm{R}}_i$ and from the MAJ to each ${\rm{R}}_i$ would become
\begin{equation}
\setcounter{equation}{7}
\begin{split}{}
{{\bf{h}}_i} =& \sqrt {\frac{\varsigma }{{\varsigma  + 1}}} {\overline {\bf{h}} _i} + \sqrt {\frac{1}{{\varsigma  + 1}}} {\widetilde {\bf{h}}_i},\\
{{\bf{g}}_i}({\bf{x}}) =& \sqrt {\frac{\varsigma }{{\varsigma  + 1}}} {\overline {\bf{g}} _i}({\bf{x}}) + \sqrt {\frac{1}{{\varsigma  + 1}}} {\widetilde {\bf{g}}_i}({\bf{x}}),
\end{split}
\end{equation}
where $\varsigma $ is the Rician factor, ${\overline {\bf{h}} _i}$ and ${\overline {\bf{g}} _i}({\bf{x}})$ are the LoS components, which have the same form as the channel vectors shown in (1) and (2), respectively. In addition, ${\widetilde {\bf{h}}_i}$ and ${\widetilde {\bf{g}}_i}({\bf{x}})$ are the Non-LoS (NLoS) components, which can be expressed as ${\widetilde {\bf{h}}_i} = {\widehat {\bf{h}}_i}/\sqrt {d_{{S_i}}^\tau }$ and ${\widetilde {\bf{g}}_i}({\bf{x}}) = {\widehat {\bf{g}}_i}({\bf{x}})/\sqrt {d_{{J_i}}^\tau } $, respectively, where ${\widehat {\bf{h}}_i},{\widehat {\bf{g}}_i}({\bf{x}}) \in {{\mathbb{C}}^{M \times 1}}$ and each entry of ${\widehat {\bf{h}}_i}$ and ${\widehat {\bf{g}}_i}({\bf{x}})$ is the independent and circular symmetric complex Gaussian random variable with zero mean and unit variance.

Considering the random characteristics of the Rician channel, for the suspicious system,
\begin{itemize}
\item[$\bullet$] Each ${\rm{R}}_i$ should send the pilot signal to the ST before each transmission slot, based on which the ST can effectively estimate the downlink channels for subsequent beamforming and power allocations.
\end{itemize}

For the legitimate party,
\begin{itemize}
\item[$\bullet$] First, unlike the ST, the MAJ cannot acquire the NLoS components of the suspicious channels.
\item[$\bullet$] Second, via detecting the pilot signal of each ${\rm{R}}_i$, the MAJ can estimate the jamming channel coefficients for optimizing its jamming beamforming.
\item[$\bullet$] Third, it is impractical for the MAJ to frequently vary its antenna positions in each transmission slot for acquiring better jamming channel conditions, since this action will cause huge energy consumptions and time delays. This implies that the MAJ should optimize its antenna positions based on the statistical CSI in advance.
\end{itemize}

Via the above analysis, from the perspective of the MAJ, the optimization objective should be minimizing the sum ergodic rate or minimum ergodic rate of the suspicious communications, which is given by
\begin{equation}
\setcounter{equation}{8}
\begin{split}{}
\mathop {\min }\limits_{\bf{x}} {{\mathbb{E}}_{\left\{ {{{\widetilde {\bf{g}}}_i}({\bf{x}})} \right\}_{i = 1}^K}}\left[ {\mathop {\min }\limits_{{{\bf{w}}_J}} {{\mathbb{E}}_{\left\{ {{{\widetilde {\bf{h}}}_i}} \right\}_{i = 1}^K}}\left[ {{R_\vartheta }(\left\{ {{P_i}} \right\},{{\bf{w}}_J},{\bf{x}})} \right]} \right],
\end{split}
\end{equation}
where $\vartheta  \in \left\{ {{\rm{sum}},\min } \right\}$. Accordingly, since ${\bf{x}}$ and ${{\bf{w}}_J}$ are optimized in the slow and fast timescale, respectively, the optimization method based on the two-timescale design in our previous work \cite{Guo_twotimescale} can be similarly adopted to solve the corresponding problem: i) first, given ${\bf{x}}$, optimize ${{\bf{w}}_J}$ to minimize the expectation of the inner layer; ii) second, optimize ${\bf{x}}$ to minimize the expectation of the outer layer.

The adaptability of our proposed scheme to the Rician channel: under the Rician channel conditions, the MAJ can still perform whole optimizations based on the statistical CSI (such as the AoAs, distance and so on) as presented in this work. Afterwards, the expectation of the sum rate or minimum rate of the suspicious communications under the Rician channel environments is considered as the evaluation indicator for examining the performance of the optimized jamming beamforming and antenna positions.

  \begin{figure*}[b!]
  \hrulefill
\setcounter{mytempeqncnt}{\value{equation}}
\setcounter{equation}{25}
\begin{equation}
\begin{split}{}
&{\Theta _{2,i}}({\bf{x}},{{\bf{x}}^{(k)}}) = {\mathop{\rm Re}\nolimits} \left( {{\bf{g}}_i^H({\bf{x}})\underbrace {{{\bf{W}}_J}{{\bf{g}}_i}({{\bf{x}}^{(k)}})}_{{{\bf{f}}_i}({{\bf{x}}^{(k)}})}} \right) = {\mathop{\rm Re}\nolimits} \left( {\sum\nolimits_{j = 1}^N {g_{i,j}^H({\bf{x}}){f_{i,j}}({{\bf{x}}^{(k)}})} } \right)\\
 =& {\rm{Re}}\left( {\sum\nolimits_{j = 1}^N {\frac{{\left| {{f_{i,j}}({{\bf{x}}^{(k)}})} \right|}}{{\sqrt {d_{{J_i}}^\tau } }}{e^{j\left( {\frac{{2\pi }}{\lambda }{x_j}\sin {\varphi _i} - {\varphi _{{f_{i,j}}({{\bf{x}}^{(k)}})}}} \right)}}} } \right) = \sum\nolimits_{j = 1}^N {\frac{{\left| {{f_{i,j}}({{\bf{x}}^{(k)}})} \right|}}{{\sqrt {d_{{J_i}}^\tau } }}\cos \left( {\frac{{2\pi }}{\lambda }{x_j}\sin {\varphi _i} - {\varphi _{{f_{i,j}}({{\bf{x}}^{(k)}})}}} \right)}  .
\end{split}
\end{equation}
\setcounter{equation}{\value{mytempeqncnt}}
\vspace{-12pt}
\end{figure*}

\section{Minimizing the Sum Rate of the Suspicious Communications}
In this section, we first consider the scenario where the sum rate maximization is the objective of the suspicious communications. Then, from the perspective of the MAJ, to solve (P1), the first step is to figure out the flexible behavior of the ST given ${{{\bf{w}}_J}}$ and ${\bf{x}}$. To this end, it is well-known that the water-filling power allocations are optimal for the ST to maximize the downlink sum rate. Hence, based on (5), given ${{{\bf{w}}_J}}$ and ${\bf{x}}$, it can be derived that
\begin{equation}
\setcounter{equation}{9}
\begin{split}{}
P_i^* = \max \left( {\eta  - \frac{{{\sigma ^2} + {Q_J}{{\left| {{\bf{w}}_J^H{{\bf{g}}_i}({\bf{x}})} \right|}^2}}}{{{{\left| {{\bf{w}}_{i,{\rm{ZF}}}^H{{\bf{h}}_i}} \right|}^2}}},0} \right),
\end{split}
\end{equation}
where $\eta $ is the water-filling level, which is the solution of the equality $\sum\nolimits_{i = 1}^K {P_i^*}  = {P_{{\rm{sum}}}}$. Substituting (9) into (6a), we have
\begin{equation}
\begin{split}{}
&{R_{{\rm{sum}}}}(\left\{ {P_i^*} \right\}_{i = 1}^K,{{\bf{w}}_J},{\bf{x}}) \buildrel \Delta \over = {R_{{\rm{sum}}}}(\eta ,{{\bf{w}}_J},{\bf{x}})\\
 =& \sum\nolimits_{i = 1}^K {{{\log }_2}\left( {\max \left( {\eta \frac{{{{\left| {{\bf{w}}_{i,{\rm{ZF}}}^H{{\bf{h}}_i}} \right|}^2}}}{{{\sigma ^2} + {Q_J}{{\left| {{\bf{w}}_J^H{{\bf{g}}_i}({\bf{x}})} \right|}^2}}},1} \right)} \right)} \\
 =& \frac{1}{{\ln 2}}\sum\nolimits_{i = 1}^K {\max \left( {\ln \left( {\eta \frac{{{{\left| {{\bf{w}}_{i,{\rm{ZF}}}^H{{\bf{h}}_i}} \right|}^2}}}{{{\sigma ^2} + {Q_J}{{\left| {{\bf{w}}_J^H{{\bf{g}}_i}({\bf{x}})} \right|}^2}}}} \right),0} \right)}.
\end{split}
\end{equation}
In addition, based on (9), the constraint in (6e) can be equivalently simplified as
\begin{equation}
\begin{split}{}
\sum\nolimits_{i = 1}^K {\max \left( {\eta  - \frac{{{\sigma ^2} + {Q_J}{{\left| {{\bf{w}}_J^H{{\bf{g}}_i}({\bf{x}})} \right|}^2}}}{{{{\left| {{\bf{w}}_{i,{\rm{ZF}}}^H{{\bf{h}}_i}} \right|}^2}}},0} \right)} \mathop  \ge \limits^{(a)} {P_{{\rm{sum}}}},
\end{split}
\end{equation}
where the inequality (a) is established mainly due to that for the MAJ, the objective is to minimize the downlink sum rate, which is monotonically increasing with respect to (w.r.t.) ${P_i^*}$, $\forall i = 1,...,K$. Therefore, for the ST, by setting $\sum\nolimits_{i = 1}^K {P_i^*}  \ge {P_{{\rm{sum}}}}$, it can improve its transmit power as far as possible to resist the interference caused by the MAJ. Otherwise, if we replace the constraint in (6e) with $\sum\nolimits_{i = 1}^K {P_i^*}  \le {P_{{\rm{sum}}}}$, then at the optimal solution, the power allocations of the ST will become $P_i^* = 0$, $\forall i = 1,...,K$. This is obviously not in line with the actual situation.

According to (10) and (11), we can simplify (P1) as
 \begin{align}
&({\rm{P2}}):{\rm{  }}\mathop {\min }\limits_{\eta ,{{\bf{w}}_J},{\bf{x}}} \ {R_{{\rm{sum}}}}(\eta ,{{\bf{w}}_J},{\bf{x}}) \tag{${\rm{12a}}$}\\
{\rm{              }}&{\rm{s.t.}} \ \quad \ (6{\rm{b}}), (6{\rm{c}}), (6{\rm{d}}), (11).\tag{${\rm{12b}}$}
 \end{align}

In (P2), the independent variables (${{\bf{w}}_J}$ and ${\bf{x}}$) and the dependent variable ($\eta$) are still coupled with each other in the objective, making (P2) highly non-convex. In the next, AO is exploited to solve (P2) in an alternating manner.

\subsection{Optimizing ${{\bf{w}}_J}$ and $\eta$ Given Fixed ${\bf{x}}$}
When ${\bf{x}}$ is fixed, (P2) can be simplified as
 \begin{align}
&({\rm{P2.1}}):{\rm{  }}\mathop {\min }\limits_{\eta ,{{\bf{w}}_J}} \ {R_{{\rm{sum}}}}(\eta ,{{\bf{w}}_J}) \tag{${\rm{13a}}$}\\
{\rm{              }}&{\rm{s.t.}} \ \quad \ (6{\rm{b}}), (11),\tag{${\rm{13b}}$}
 \end{align}
where ${R_{{\rm{sum}}}}(\eta ,{{\bf{w}}_J})$ is expanded as in (14). Observing (P2.1), although the constraint in (6b) is convex w.r.t. ${{\bf{w}}_J}$, the objective and the constraint in (11) are still non-convex w.r.t. ${{\bf{w}}_J}$ and $\eta $. To this end, the SCA will be exploited to iteratively update these two variables until the objective of (P2.1) converges to a stationary value. Specifically,

1) For the objective, given the known ${\eta ^{(k)}}$ and ${\bf{w}}_J^{(k)}$ in the $k$-th iteration, ${\Theta _1}(\eta )$ is upper-bounded by
\begin{equation}
\setcounter{equation}{15}
\begin{split}{}
{\Theta _1}(\eta ) \le& {\Theta _1}(\eta ,{\eta ^{(k)}}) = \ln {\eta ^{(k)}} + \frac{1}{{{\eta ^{(k)}}}}(\eta  - {\eta ^{(k)}}),
\end{split}
\end{equation}
and ${{\Theta _{2,i}}({{\bf{w}}_J})}$ is lower-bounded by \cite{MWYY_TWC}
\begin{equation}
\begin{split}{}
&{\Theta _{2,i}}({{\bf{w}}_J}) = {\bf{w}}_J^H{{\bf{G}}_i}({\bf{x}}){{\bf{w}}_J}\\
 \ge &{\Theta _{2,i}}({{\bf{w}}_J},{\bf{w}}_J^{(k)}) = 2{\rm{Re}}\left( {{\bf{w}}_J^H{{\bf{G}}_i}({\bf{x}}){\bf{w}}_J^{(k)}} \right)\\
{\rm{  }}& - {({\bf{w}}_J^{(k)})^H}{{\bf{G}}_i}({\bf{x}}){\bf{w}}_J^{(k)},
\end{split}
\end{equation}
with ${{\bf{G}}_i}({\bf{x}}) = {{\bf{g}}_i}({\bf{x}}){\bf{g}}_i^H({\bf{x}})$.

Based on (15) and (16), we now can obtain the upper bound of the objective ${R_{{\rm{sum}}}}(\eta ,{{\bf{w}}_J})$, i.e.,
\begin{equation}
\begin{split}{}
&{R_{{\rm{sum}}}}(\eta ,{{\bf{w}}_J};{\eta ^{(k)}},{\bf{w}}_J^{(k)}) = \frac{1}{{\ln 2}}\\
 \times &\sum\limits_{i = 1}^K {\max \left( {\begin{array}{*{20}{l}}
{{{\rm{C}}_{1,i}} + {\Theta _1}(\eta ,{\eta ^{(k)}})}\\
{ - \ln \left( {{\sigma ^2} + {Q_J}{\Theta _{2,i}}({{\bf{w}}_J},{\bf{w}}_J^{(k)})} \right),0}
\end{array}} \right)} ,
\end{split}
\end{equation}
which is convex w.r.t. ${{\bf{w}}_J}$ and $\eta $.

2) On the other hand, focusing on the constraint in (11), since $\max (x,0)$ is convex w.r.t. $x$ and the lower bound of $\max (x,0)$ is very hard to be derived given the known ${x^{(k)}}$, we need to find a tight approximation with clear structures to replace the left term in (11). In detail, by exploiting the log-sum-exp function \cite{jamming_relay}, we can obtain an approximation for $\max (x,0)$ as
\begin{equation}
\begin{split}{}
\max (x,0) \approx \frac{1}{\chi }\ln \left( {{e^{\chi x}} + 1} \right),
\end{split}
\end{equation}
where $\chi $ is a positive parameter. The above approximation becomes tight as $\chi  \to \infty $. In practice, $\chi$ can be set in a small range, e.g., $[4,8]$, to achieve a high approximation accuracy. Based on (18), the constraint in (11) can be replaced with
\begin{equation}
\begin{split}{}
\sum\nolimits_{i = 1}^K {\frac{1}{\chi }\ln \left( {{e^{\chi \left( {\eta   - \frac{{{\sigma ^2} + {Q_J}{{\left| {{\bf{w}}_J^H{{\bf{g}}_i}({\bf{x}})} \right|}^2}}}{{{{\left| {{\bf{w}}_{i,{\rm{ZF}}}^H{{\bf{h}}_i}} \right|}^2}}}} \right)}} + 1} \right)}  \ge {P_{{\rm{sum}}}}.
\end{split}
\end{equation}

Unfortunately, the constraint in (19) is still non-convex. To tackle it, we introduce $K$ slack variables $\left\{ {{\Psi _i}} \right\}_{i = 1}^K$ and equivalently transform the constraint in (19) as
\begin{equation}
\begin{split}{}
\left\{ {\begin{array}{*{20}{c}}
{\sum\nolimits_{i = 1}^K {\frac{1}{\chi }\ln \left( {1 + {\Psi _i}} \right)}  \ge {P_{{\rm{sum}}}}}\\
{{e^{\chi \left( {\eta   - \frac{{{\sigma ^2} + {Q_J}{{\left| {{\bf{w}}_J^H{{\bf{g}}_i}({\bf{x}})} \right|}^2}}}{{{{\left| {{\bf{w}}_{i,{\rm{ZF}}}^H{{\bf{h}}_i}} \right|}^2}}}} \right)}} \ge {\Psi _i},i = 1,...,K},
\end{array}} \right.
\end{split}
\end{equation}
which can be further simplified as
\begin{equation}
\begin{split}{}
\left\{ {\begin{array}{*{20}{c}}
{\sum\nolimits_{i = 1}^K {\frac{1}{\chi }\ln \left( {1 + {\Psi _i}} \right)}  \ge {P_{{\rm{sum}}}}}\\
{\chi \left( {\eta   - \frac{{{\sigma ^2} + {Q_J}{{\left| {{\bf{w}}_J^H{{\bf{g}}_i}({\bf{x}})} \right|}^2}}}{{{{\left| {{\bf{w}}_{i,{\rm{ZF}}}^H{{\bf{h}}_i}} \right|}^2}}}} \right) \ge \ln {\Psi _i},i = 1,...,K}.
\end{array}} \right.
\end{split}
\end{equation}

Note that the first constraint in (21) is convex w.r.t. $\left\{ {{\Psi _i}} \right\}_{i = 1}^K$. For the right term of the second constraint in (21), given the known $\Psi _i^{(k)}$ in the $k$-th iteration, we need to derive the upper bound of ${\ln {\Psi _i}}$ as $\ln {\Psi _i} \le {\Theta _3}({\Psi _i},\Psi _i^{(k)}) = \ln \Psi _i^{(k)} + \frac{1}{{\Psi _i^{(k)}}}({\Psi _i} - \Psi _i^{(k)})$, based on which the constraint in (21) can be approximated as
\begin{equation}
\begin{split}{}
\left\{ {\begin{array}{*{20}{c}}
{\sum\nolimits_{i = 1}^K {\frac{1}{\chi }\ln \left( {1 + {\Psi _i}} \right)}  \ge {P_{{\rm{sum}}}}}\\
{\chi \left( {\eta   - \frac{{{\sigma ^2} + {Q_J}{{\left| {{\bf{w}}_J^H{{\bf{g}}_i}({\bf{x}})} \right|}^2}}}{{{{\left| {{\bf{w}}_{i,{\rm{ZF}}}^H{{\bf{h}}_i}} \right|}^2}}}} \right) \ge {\Theta _3}({\Psi _i},\Psi _i^{(k)}),i = 1,...,K},
\end{array}} \right.
\end{split}
\end{equation}
which now is convex w.r.t. $\left\{ {{\Psi _i}} \right\}_{i = 1}^K$, $\eta $ and ${{\bf{w}}_J}$.

Armed with the above analysis, (P2.1) can be transformed to the following approximate problem
 \begin{align}
&({\rm{P2.2}}):{\rm{  }}\mathop {\min }\limits_{\eta ,{{\bf{w}}_J},\left\{ {{\Psi _i}} \right\}_{i = 1}^K} \ {R_{{\rm{sum}}}}(\eta ,{{\bf{w}}_J};{\eta ^{(k)}},{\bf{w}}_J^{(k)}) \tag{${\rm{23a}}$}\\
{\rm{              }}& \ {\rm{s.t.}} \ \quad \ \ \ (6{\rm{b}}), (22).\tag{${\rm{23b}}$}
 \end{align}
Note that (P2.2) is a convex optimization problem, which thus can be iteratively solved by using existing solvers, e.g. CVX.

Complexity analysis: There are $K + 1$ real variables $\eta $ and $\left\{ {{\Psi _i}} \right\}_{i = 1}^K$, and one $N \times 1$ complex variable ${{\bf{w}}_J}$ in (P2.2). Therefore, the complexity of solving (P2.2) is about ${\cal O}\left( {{I_{{\rm{iter}},1}}{{(K + 1 + 2N)}^{3.5}}} \right)$, where ${{I_{{\rm{iter}},1}}}$ is the number of iterations for satisfying the convergence condition.

\subsection{Optimizing ${\bf{x}}$ and $\eta$ Given Fixed ${{\bf{w}}_J}$}
When ${{\bf{w}}_J}$ is fixed, (P2) can be simplified as
 \begin{align}
&({\rm{P2.3}}):{\rm{  }}\mathop {\min }\limits_{\eta ,{\bf{x}}} \ {R_{{\rm{sum}}}}(\eta ,{\bf{x}}) \tag{${\rm{24a}}$}\\
{\rm{              }}&\ \ {\rm{s.t.}} \ \quad \ (6{\rm{c}}), (6{\rm{d}}), (11),\tag{${\rm{24b}}$}
 \end{align}
where ${R_{{\rm{sum}}}}(\eta ,{\bf{x}})$ equals ${R_{{\rm{sum}}}}(\eta ,{{\bf{w}}_J})$ shown in (14), by just redefining ${{\Theta _{2,i}}({{\bf{w}}_J})}$ as ${{\Theta _{2,i}}({\bf{x}}) = {{\left| {{\bf{w}}_J^H{{\bf{g}}_i}({\bf{x}})} \right|}^2}}$. For (P2.3), note that the constraints in (6c) and (6d) are convex w.r.t. ${\bf{x}}$, but the objective and the constraint in (11) are non-convex w.r.t. ${\bf{x}}$ and $\eta $. Similarly, the SCA should be employed to iteratively solve (P2.3). Specifically,

1) Focusing on the objective of (P2.3), we now ignore the same step for deriving the upper bound of ${\Theta _1}(\eta )$ shown in (15), and just aim to derive the lower bound of ${{\Theta _{2,i}}({\bf{x}})}$, $\forall i = 1,...,K$. Specifically, we can first obtain
\begin{equation}
\setcounter{equation}{25}
\begin{split}{}
&{\Theta _{2,i}}({\bf{x}}) = {\bf{g}}_i^H({\bf{x}}){{\bf{W}}_J}{{\bf{g}}_i}({\bf{x}})\\
 \ge &2\underbrace {{\rm{Re}}\left( {{\bf{g}}_i^H({\bf{x}}){{\bf{W}}_J}{{\bf{g}}_i}({{\bf{x}}^{(k)}})} \right)}_{{\Theta _{2,i}}({\bf{x}},{{\bf{x}}^{(k)}})} - \underbrace {{\bf{g}}_i^H({{\bf{x}}^{(k)}}){{\bf{W}}_J}{{\bf{g}}_i}({{\bf{x}}^{(k)}})}_{{{\rm{C}}_{2,i}}},
\end{split}
\end{equation}
where ${{\bf{W}}_J} = {{\bf{w}}_J}{\bf{w}}_J^H$. However, the term ${{\Theta _{2,i}}({\bf{x}},{{\bf{x}}^{(k)}})}$ in (25) is still non-convex w.r.t. ${\bf{x}}$. To proceed, we now expand ${{\Theta _{2,i}}({\bf{x}},{{\bf{x}}^{(k)}})}$ as in (26), where ${g_{i,j}}({\bf{x}})$ and ${{f_{i,j}}({{\bf{x}}^{(k)}})}$ are the $j$-th element in ${{\bf{g}}_i}({\bf{x}})$ and the $j$-th element in ${{{\bf{f}}_i}({{\bf{x}}^{(k)}})}$, respectively, and ${{\varphi _{{f_{i,j}}({{\bf{x}}^{(k)}})}}}$ is the phase of ${{f_{i,j}}({{\bf{x}}^{(k)}})}$. Subsequently, by resorting to the second-order Taylor expansion \cite{MWYY_TWC}, given $x_j^{(k)}$, we can derive
\begin{equation}
\setcounter{equation}{27}
\begin{split}{}
&\cos \left( {\frac{{2\pi }}{\lambda }{x_j}\sin {\varphi _i} - {\varphi _{{f_{i,j}}({{\bf{x}}^{(k)}})}}} \right)\\
\ge & {\Theta _4}({x_j},x_j^{(k)}) = \cos \left( {\frac{{2\pi }}{\lambda }x_j^{(k)}\sin {\varphi _i} - {\varphi _{{f_{i,j}}({{\bf{x}}^{(k)}})}}} \right) \\
 -& \frac{{2\pi }}{\lambda }\sin {\varphi _i}\sin \left( {\frac{{2\pi }}{\lambda }x_j^{(k)}\sin {\varphi _i} - {\varphi _{{f_{i,j}}({{\bf{x}}^{(k)}})}}} \right)\\
\times & \left( {{x_j} - x_j^{(k)}} \right) - \frac{{\delta _i}}{2}{\left( {{x_j} - x_j^{(k)}} \right)^2},
\end{split}
\end{equation}
with ${\delta _i} = {\left( {\frac{{2\pi }}{\lambda }\sin {\varphi _i}} \right)^2}$. Then, based on (26) and (27), the lower bound of ${\Theta _{2,i}}({\bf{x}})$ can be given by
\begin{equation}
\begin{split}{}
&{\Theta _{2,i}}({\bf{x}}) \ge {\Theta _{5,i}}({\bf{x}},{{\bf{x}}^{(k)}}) \\
  =& 2\sum\nolimits_{j = 1}^N {\frac{{\left| {{f_{i,j}}({{\bf{x}}^{(k)}})} \right|}}{{\sqrt {d_{{J_i}}^\tau } }}} {\Theta _4}({x_j},x_j^{(k)}) - {{\rm{C}}_{2,i}}.
\end{split}
\end{equation}

Based on (15) and (28), we can obtain the upper bound of the objective ${R_{{\rm{sum}}}}(\eta ,{\bf{x}})$ as
\begin{equation}
\begin{split}{}
&{R_{{\rm{sum}}}}(\eta ,{\bf{x}};{\eta ^{(k)}},{{\bf{x}}^{(k)}}) = \frac{1}{{\ln 2}}\\
 \times &\sum\nolimits_{i = 1}^K {\max \left( \begin{array}{l}
{{\rm{C}}_{1,i}} + {\Theta _1}(\eta ,{\eta ^{(k)}})\\
 - \ln \left( {{\sigma ^2} + {Q_J}{\Theta _{5,i}}({\bf{x}},{{\bf{x}}^{(k)}})} \right),0
\end{array} \right)}.
\end{split}
\end{equation}
which is convex w.r.t. ${{\bf{x}}}$ and $\eta $.

2) For the non-convex constraint in (11), via the above analysis we have obtained an approximate alternative as shown in (21). However, unlike the case in subsection A, the left term of the second constraint in (21) is not convex w.r.t. ${{\bf{x}}}$. This motivates us to further derive its lower bound, which is equivalent to finding the upper bound of the term ${{{\left| {{\bf{w}}_J^H{{\bf{g}}_i}({\bf{x}})} \right|}^2}}$, $i = 1,...,K$. Specifically, based on \cite{Cunhua}, we have
\begin{equation}
\begin{split}{}
&{\left| {{\bf{w}}_J^H{{\bf{g}}_i}({\bf{x}})} \right|^2} = {\bf{g}}_i^H({\bf{x}}){{\bf{W}}_J}{{\bf{g}}_i}({\bf{x}})\\
 \le& {\bf{g}}_i^H({\bf{x}}){{\bf{\Phi }}_J}{{\bf{g}}_i}({\bf{x}}) + \underbrace {{\bf{g}}_i^H({{\bf{x}}^{(k)}})({{\bf{\Phi }}_J} - {{\bf{W}}_J}){{\bf{g}}_i}({{\bf{x}}^{(k)}})}_{{{\rm{C}}_{3,i}}}\\
 &- 2{\mathop{\rm Re}\nolimits} \left( {{\bf{g}}_i^H({\bf{x}})\underbrace {({{\bf{\Phi }}_J} - {{\bf{W}}_J}){{\bf{g}}_i}({{\bf{x}}^{(k)}})}_{{{\bf{v}}_i}({{\bf{x}}^{(k)}})}} \right),
\end{split}
\end{equation}
where ${{\bf{\Phi }}_J} = {\xi _{\max }}({{\bf{W}}_J}){{\bf{I}}_N}$ and ${\xi _{\max }}({{\bf{W}}_J})$ is the principal eigenvalue of ${{\bf{W}}_J}$. Then, focusing on (30), since ${\bf{g}}_i^H({\bf{x}}){{\bf{g}}_i}({\bf{x}}) = N/d_{{J_i}}^\tau $, it can be derived that ${\bf{g}}_i^H({\bf{x}}){{\bf{\Phi }}_J}{{\bf{g}}_i}({\bf{x}}) = {\xi _{\max }}({{\bf{W}}_J})N/d_{{J_i}}^\tau $, which is not related to ${\bf{x}}$. Further, based on (26)$-$(28), the lower bound of ${\mathop{\rm Re}\nolimits} \left( {{\bf{g}}_i^H({\bf{x}}){{\bf{v}}_i}({{\bf{x}}^{(k)}})} \right)$ can be similarly derived as
\begin{equation}
\begin{split}{}
&{\mathop{\rm Re}\nolimits} \left( {{\bf{g}}_i^H({\bf{x}}){{\bf{v}}_i}({{\bf{x}}^{(k)}})} \right)\\
 \ge &{\Theta _{7,i}}({\bf{x}},{{\bf{x}}^{(k)}}) = \sum\nolimits_{j = 1}^N {\frac{{\left| {{v_{i,j}}({{\bf{x}}^{(k)}})} \right|}}{{\sqrt {d_{{J_i}}^\tau } }}} {\Theta _6}({x_j},x_j^{(k)}),
\end{split}
\end{equation}
where ${{v_{i,j}}({{\bf{x}}^{(k)}})}$ is the $j$-th element in ${{{\bf{v}}_i}({{\bf{x}}^{(k)}})}$ and
\begin{equation}
\begin{split}{}
&{\Theta _6}({x_j},x_j^{(k)}) \\
=& \cos \left( {\frac{{2\pi }}{\lambda }x_j^{(k)}\sin {\varphi _i} - {\varphi _{{v_{i,j}}({{\bf{x}}^{(k)}})}}} \right)\\
-& \frac{{2\pi }}{\lambda }\sin {\varphi _i}\sin \left( {\frac{{2\pi }}{\lambda }x_j^{(k)}\sin {\varphi _i} - {\varphi _{{v_{i,j}}({{\bf{x}}^{(k)}})}}} \right)\\
& \times \left( {{x_j} - x_j^{(k)}} \right) - \frac{{\delta _i} }{2}{\left( {{x_j} - x_j^{(k)}} \right)^2},
\end{split}
\end{equation}
with ${{\varphi _{{v_{i,j}}({{\bf{x}}^{(k)}})}}}$ denoting the phase of ${{v_{i,j}}({{\bf{x}}^{(k)}})}$.

Based on (31), the upper bound of ${\left| {{\bf{w}}_J^H{{\bf{g}}_i}({\bf{x}})} \right|^2}$ can be derived as
\begin{equation}
\begin{split}{}
&{\left| {{\bf{w}}_J^H{{\bf{g}}_i}({\bf{x}})} \right|^2} \le {\Theta _{8,i}}({\bf{x}},{{\bf{x}}^{(k)}})\\
 =& {\xi _{\max }}({{\bf{W}}_J})N/d_{{J_i}}^\tau  + {{\rm{C}}_{3,i}} - 2{\Theta _{7,i}}({\bf{x}},{{\bf{x}}^{(k)}}).
\end{split}
\end{equation}

Then, focusing on (22), replacing the term ${\left| {{\bf{w}}_J^H{{\bf{g}}_i}({\bf{x}})} \right|^2}$ in the second constraint with ${\Theta _{8,i}}({\bf{x}},{{\bf{x}}^{(k)}})$, we can obtain that
\begin{equation}
\begin{split}{}
\left\{ {\begin{array}{*{20}{c}}
{\sum\nolimits_{i = 1}^K {\frac{1}{\chi }\ln \left( {1 + {\Psi _i}} \right)}  \ge {P_{{\rm{sum}}}}}\\
{\chi \left( {\eta  - \frac{{{\sigma ^2} + {Q_J}{\Theta _{8,i}}({\bf{x}},{{\bf{x}}^{(k)}})}}{{{{\left| {{\bf{w}}_{i,{\rm{ZF}}}^H{{\bf{h}}_i}} \right|}^2}}}} \right) \ge {\Theta _3}({\Psi _i},\Psi _i^{(k)}),i = 1,...,K,}
\end{array}} \right.
\end{split}
\end{equation}
which is convex w.r.t. $\left\{ {{\Psi _i}} \right\}_{i = 1}^K$, $\eta $ and ${{\bf{x}}}$.

Armed with the above analysis, (P2.3) now can be transformed to the following approximate problem
 \begin{align}
&({\rm{P2.4}}):{\rm{  }}\mathop {\min }\limits_{\eta ,{{\bf{x}}},\left\{ {{\Psi _i}} \right\}_{i = 1}^K} \ {R_{{\rm{sum}}}}(\eta ,{{\bf{x}}};{\eta ^{(k)}},{\bf{x}}^{(k)}) \tag{${\rm{35a}}$}\\
{\rm{              }}& \ {\rm{s.t.}} \ \quad \ \ \ (6{\rm{c}}), (6{\rm{d}}), (34).\tag{${\rm{35b}}$}
 \end{align}
Note that (P2.4) is a convex optimization problem, which can be iteratively solved by using CVX.

Complexity analysis: Similarly, there are $K + 1 +N$ real variables in (P2.4). Therefore, the complexity of solving (P2.4) is about ${\cal O}\left( {{I_{{\rm{iter}},2}}{{(K + 1 + N)}^{3.5}}} \right)$, where ${{I_{{\rm{iter}},2}}}$ is the number of iterations for satisfying the convergence condition.

\begin{algorithm}
\caption{The AO for Solving (P2)}
  \begin{algorithmic}[1]

\State \textbf{{Input:}} ${\eta ^{(1)}}$, ${\bf{w}}_J^{(1)}$, $\left\{ {\Psi _i^{(1)}} \right\}_{i = 1}^K$, ${{\bf{x}}^{(1)}}$, and $k = 1$.

\State \textbf{Repeat:}


\State \quad \textbf{Repeat:}

\State \quad Solve (P2.2) and output $\eta $, ${{\bf{w}}_J}$ and $\left\{ {{\Psi _i}} \right\}_{i = 1}^K$. 

\State \quad $k \leftarrow k + 1$.

\State \quad Set ${\eta ^{(k)}} = \eta $, ${\bf{w}}_J^{(k)} = {{\bf{w}}_J}$ and $\left\{ {\Psi _i^{(k)} = {\Psi _i}} \right\}_{i = 1}^K$.

\State \quad \textbf{Until:} ${R_{{\rm{sum}}}}(\eta ,{{\bf{w}}_J};{\eta ^{(k - 1)}},{\bf{w}}_J^{(k - 1)})$ converges to a stationary value.



\State \quad \textbf{Repeat:}

\State \quad Solve (P2.4) and output $\eta $, ${{\bf{x}}}$ and $\left\{ {{\Psi _i}} \right\}_{i = 1}^K$. 

\State \quad $k \leftarrow k + 1$;

\State \quad Set ${\eta ^{(k)}} = \eta $, ${\bf{x}}^{(k)} = {{\bf{x}}}$ and $\left\{ {\Psi _i^{(k)} = {\Psi _i}} \right\}_{i = 1}^K$.

\State \quad \textbf{Until:} ${R_{{\rm{sum}}}}(\eta ,{{\bf{x}}};{\eta ^{(k - 1)}},{\bf{x}}^{(k - 1)})$ converges to a stationary value.


\State \textbf{Until:} $\left| \begin{array}{l}
{R_{{\rm{sum}}}}(\eta ,{\bf{x}};{\eta ^{(k - 1)}},{{\bf{x}}^{(k - 1)}})\\
 - {R_{{\rm{sum}}}}(\eta ,{{\bf{w}}_J};{\eta ^{(k - 1)}},{\bf{w}}_J^{(k - 1)})
\end{array} \right| \le \varepsilon $.

\State \textbf{Output:} ${{\bf{w}}_J}$ and ${\bf{x}}$.
  \end{algorithmic}
\end{algorithm}

\subsection{Overall Algorithm, Complexity and Convergence}
We now summarize our proposed AO algorithm for solving (P2), as shown in Algorithm 1. The initial input includes the known ${\eta ^{(1)}}$, ${\bf{w}}_J^{(1)}$, $\left\{ {\Psi _i^{(1)}} \right\}_{i = 1}^K$ and ${{\bf{x}}^{(1)}}$. Afterwards, in each outer iteration, (P2.2) and (P2.4) are solved sequentially by implementing steps 3$-$7 and steps 8$-$12 in Algorithm 1, respectively. Finally, the outer iteration can be terminated if the condition in step 13 is satisfied, where $\varepsilon $ is the convergence accuracy.
%
%
%

In addition, via the above analysis, we can conclude that the total complexity of Algorithm 1 is about ${\cal O}\left( {{I_{{\rm{iter}},{\rm{out}}}}\left( {{I_{{\rm{iter}},1}}{{(K + 1 + 2N)}^{3.5}} + {I_{{\rm{iter}},2}}{{(K + 1 + N)}^{3.5}}} \right)} \right)$, where ${{I_{{\rm{iter}},{\rm{out}}}}}$ is the number of outer iterations. Furthermore, note that
\begin{equation}
\setcounter{equation}{36}
\begin{split}{}
&{R_{{\rm{sum}}}}(\eta ,{{\bf{w}}_{J,m - 1}},{{\bf{x}}_{m - 1}})\\
\mathop  \ge \limits^{(b)} &{R_{{\rm{sum}}}}(\eta ,{{\bf{w}}_{J,m}},{{\bf{x}}_{m - 1}})\\
\mathop  \ge \limits^{(c)} &{R_{{\rm{sum}}}}(\eta ,{{\bf{w}}_{J,m}},{{\bf{x}}_m}),
\end{split}
\end{equation}
where in the $m$-th outer iteration, the inequality (b) is established after running (P2.2) and the inequality (c) is established after running (P2.4), and ${{\bf{w}}_{J,m}}$ and ${{\bf{x}}_{m}}$ denote the outputs of (P2.2) and (P2.4) in the corresponding outer iteration, respectively. In addition, since the objective shown in (12a) has the lower bound which is not smaller than zero, our proposed algorithm can strictly converge to a stationary point.

\subsection{Initialization}
In this subsection, we further provide the detailed initialization scheme for Algorithm 1. Specifically, during the initialization phase: i) the positions of MAs at the MAJ are set as ${{\bf{x}}^{(1)}} = \left[ {0:L/(N - 1):L} \right]$. This configuration could avoid the coupling effect between the adjacent MAs as far as possible and concurrently allow each antenna to move flexibly within its adjacent space to the maximum extent; ii) the jamming beamforming is set as the normalized sum of all jamming channel vectors \cite{YANG_GANG}, i.e., ${\bf{w}}_J^{(1)} = \sum\nolimits_{i = 1}^K {{{\bf{g}}_i}({{\bf{x}}^{(1)}})} /\left\| {\sum\nolimits_{i = 1}^K {{{\bf{g}}_i}({{\bf{x}}^{(1)}})} } \right\|$; iii) the water-filling level is set as ${\eta ^{(1)}} = 5$ and other slack variables are set as $\Psi _i^{(1)} = 100,\forall i = 1,...,K$. These values are empirically determined through comparative experiments.

\section{Minimizing the Fairness of the Suspicious Communications}
In this section, we consider another scenario where the minimum rate maximization is the objective of the suspicious communications. Then, from the perspective of the MAJ, to solve (P1), as a preliminary it should also figure out the optimal power allocations of the ST given ${{{\bf{w}}_J}}$ and ${\bf{x}}$. To this end, based on (4), it is not difficult to verify that $\left\{ {P_i^*} \right\}$ are the solutions of the following equations
\begin{equation}
\begin{split}{}
\left\{ {\begin{array}{*{20}{c}}
{\frac{{{P_1}{{\left| {{\bf{w}}_{1,{\rm{ZF}}}^H{{\bf{h}}_1}} \right|}^2}}}{{{Q_J}{{\left| {{\bf{w}}_J^H{{\bf{g}}_1}({\bf{x}})} \right|}^2} + {\sigma ^2}}} = ... = \frac{{{P_K}{{\left| {{\bf{w}}_{K,{\rm{ZF}}}^H{{\bf{h}}_K}} \right|}^2}}}{{{Q_J}{{\left| {{\bf{w}}_J^H{{\bf{g}}_K}({\bf{x}})} \right|}^2} + {\sigma ^2}}}}\\
{\sum\nolimits_{i = 1}^K {{P_i}}  = {P_{{\rm{sum}}}}},
\end{array}} \right.
\end{split}
\end{equation}
solving which returns that
\begin{equation}
\begin{split}{}
P_i^* = \frac{{{P_{{\rm{sum}}}}}}{{\sum\nolimits_{j = 1}^K {\frac{{{{\left| {{\bf{w}}_{i,{\rm{ZF}}}^H{{\bf{h}}_i}} \right|}^2}\left( {{Q_J}{{\left| {{\bf{w}}_J^H{{\bf{g}}_j}({\bf{x}})} \right|}^2} + {\sigma ^2}} \right)}}{{{{\left| {{\bf{w}}_{j,{\rm{ZF}}}^H{{\bf{h}}_j}} \right|}^2}\left( {{Q_J}{{\left| {{\bf{w}}_J^H{{\bf{g}}_i}({\bf{x}})} \right|}^2} + {\sigma ^2}} \right)}}} }},i = 1,...,K.
\end{split}
\end{equation}
Then, substituting $P_i^*$ into (4), we can derive the exact expression of each ${{\gamma _i}(P_i^*,{{\bf{w}}_J},{\bf{x}})}$ as
\begin{equation}
\begin{split}{}
{\gamma _i}(P_i^*,{{\bf{w}}_J},{\bf{x}}) = \frac{{{P_{{\rm{sum}}}}}}{{\sum\nolimits_{j = 1}^K {\frac{{{Q_J}{{\left| {{\bf{w}}_J^H{{\bf{g}}_j}({\bf{x}})} \right|}^2} + {\sigma ^2}}}{{{{\left| {{\bf{w}}_{j,{\rm{ZF}}}^H{{\bf{h}}_j}} \right|}^2}}}} }}, \forall i = 1,...,K.
\end{split}
\end{equation}
Based on (39), ${\rm{Benefit}}\left( {\left\{ {P_i^*} \right\},{{\bf{w}}_J},{\bf{x}}} \right)$ can be derived as
\begin{equation}
\begin{split}{}
&{\rm{Benefit}}\left( {\left\{ {P_i^*} \right\},{{\bf{w}}_J},{\bf{x}}} \right)\\
 =& \mathop {\min }\limits_i \left\{ {{{\log }_2}\left( {1 + {\gamma _i}(P_i^*,{{\bf{w}}_J},{\bf{x}})} \right)} \right\}\\
 =& {\log _2}\left( {1 + \frac{{{P_{{\rm{sum}}}}}}{{\sum\nolimits_{j = 1}^K {\frac{{{Q_J}{{\left| {{\bf{w}}_J^H{{\bf{g}}_j}({\bf{x}})} \right|}^2} + {\sigma ^2}}}{{{{\left| {{\bf{w}}_{j,{\rm{ZF}}}^H{{\bf{h}}_j}} \right|}^2}}}} }}} \right).
\end{split}
\end{equation}
Then, since the closed-form $P_i^*$, $i = 1,...,K$, has been obtained and substituted into the objective, the constraint in (6e) can be deleted. Further, based on the fact that minimizing ${\log _2}\left( {1 + \frac{1}{x + c}} \right)$ ($c$ is a constant) is equivalent to maximizing $x$, we can simplify (P1) as
 \begin{align}
&({\rm{P3}}):{\rm{  }}\mathop {\max }\limits_{{{{\bf{w}}_J},{\bf{x}}}} \ \sum\nolimits_{j = 1}^K {\frac{{{Q_J}{{\left| {{\bf{w}}_J^H{{\bf{g}}_j}({\bf{x}})} \right|}^2}}}{{{{\left| {{\bf{w}}_{j,{\rm{ZF}}}^H{{\bf{h}}_j}} \right|}^2}}}}  \tag{${\rm{41a}}$}\\
{\rm{              }}&{\rm{s.t.}} \quad (6{\rm{b}}), (6{\rm{c}}), (6{\rm{d}}). \tag{${\rm{41b}}$}
 \end{align}

Focusing on (P3), we can observe that the variables ${{\bf{w}}_J}$ and ${\bf{x}}$ are coupled with each other in the objective, which results in the highly non-convexity of (P3). To handle this difficulty, we re-express the objective of (P3) as
\begin{equation}
\setcounter{equation}{42}
\begin{split}{}
&\sum\nolimits_{j = 1}^K {\frac{{{Q_J}{{\left| {{\bf{w}}_J^H{{\bf{g}}_j}({\bf{x}})} \right|}^2}}}{{{{\left| {{\bf{w}}_{j,{\rm{ZF}}}^H{{\bf{h}}_j}} \right|}^2}}}} \\
 =& \sum\nolimits_{j = 1}^K {{\rm{Tr}}\left( {{{\rm{D}}_{1,j}}{{\bf{g}}_j}({\bf{x}}){\bf{g}}_j^H({\bf{x}}){{\bf{W}}_J}} \right)} \\
 =& {\rm{Tr}}\left( {{\bf{G}}({\bf{x}}){{\bf{G}}^H}({\bf{x}}){{\bf{W}}_J}} \right),
\end{split}
\end{equation}
where ${{\rm{D}}_{1,j}} = {Q_J}/{\left| {{\bf{w}}_{j,{\rm{ZF}}}^H{{\bf{h}}_j}} \right|^2}$, ${{\bf{W}}_J}$ has been defined in (25) and ${\bf{G}}({\bf{x}}) = \left[ {\begin{array}{*{20}{c}}
{\sqrt {{{\rm{D}}_{1,1}}} {{\bf{g}}_1}({\bf{x}})}&{...}&{\sqrt {{{\rm{D}}_{1,K}}} {{\bf{g}}_K}({\bf{x}})}
\end{array}} \right] \in {{\mathbb{C}}^{N \times K}}$. Based on (42), given ${\bf{x}}$, the optimal ${{\bf{W}}_J}$ subject to the constraint in (6b) has the closed-form solution as \cite{MIMO_BROAD}
\begin{equation}
{\bf{W}}_J^* = {{\bf{u}}_{\max }}\left( {{\bf{G}}({\bf{x}}){{\bf{G}}^H}({\bf{x}})} \right){\bf{u}}_{\max }^H\left( {{\bf{G}}({\bf{x}}){{\bf{G}}^H}({\bf{x}})} \right),
\end{equation}
where ${{\bf{u}}_{\max }}\left( {{\bf{G}}({\bf{x}}){{\bf{G}}^H}({\bf{x}})} \right)$ is the principal eigenvector of ${{\bf{G}}({\bf{x}}){{\bf{G}}^H}({\bf{x}})}$. Substituting ${\bf{W}}_J^*$ into (42), the objective of (P3) can be simplified as
\begin{equation}
{\rm{Tr}}\left( {{\bf{G}}({\bf{x}}){{\bf{G}}^H}({\bf{x}}){\bf{W}}_J^*} \right) = {\xi _{\max }}\left( {{\bf{G}}({\bf{x}}){{\bf{G}}^H}({\bf{x}})} \right).
\end{equation}
where ${\xi _{\max }}\left( {{\bf{G}}({\bf{x}}){{\bf{G}}^H}({\bf{x}})} \right)$ is the principal eigenvalue of ${{\bf{G}}({\bf{x}}){{\bf{G}}^H}({\bf{x}})}$. Based on (44), (P3) can be simplified as a problem related only to ${\bf{x}}$, i.e.,
 \begin{align}
&({\rm{P3.1}}):{\rm{  }}\mathop {\max }\limits_{{\bf{x}}} \ {\xi _{\max }}\left( {{\bf{G}}({\bf{x}}){{\bf{G}}^H}({\bf{x}})} \right) \tag{${\rm{45a}}$}\\
{\rm{              }}&\ {\rm{s.t.}} \ \quad \ (6{\rm{c}}), (6{\rm{d}}).\tag{${\rm{45b}}$}
 \end{align}

Although the objective of (P3.1) is only related to ${\bf{x}}$, its structure is much complex which hinders the subsequent analysis. To tackle this difficulty, we now provide an equivalent expression for ${\xi _{\max }}\left( {{\bf{G}}({\bf{x}}){{\bf{G}}^H}({\bf{x}})} \right)$, i.e.,
\begin{equation}
\setcounter{equation}{46}
\begin{split}{}
{\xi _{\max }}\left( {{\bf{G}}({\bf{x}}){{\bf{G}}^H}({\bf{x}})} \right) =& {\xi _{\max }}\left( {{{\bf{G}}^H}({\bf{x}}){\bf{G}}({\bf{x}})} \right)\\
 =& \mathop {\max }\limits_{{{\left\| {\bf{y}} \right\|}^2} = 1} {{\bf{y}}^H}{{\bf{G}}^H}({\bf{x}}){\bf{G}}({\bf{x}}){\bf{y}},
 \end{split}
\end{equation}
where ${\bf{y}} \in {{\mathbb{C}}^{K \times 1}}$ is a complex vector to be optimized. Note that in (46), we focus on the principal eigenvalue of ${{\bf{G}}^H}({\bf{x}}){\bf{G}}({\bf{x}}) \in {{\mathbb{C}}^{K \times K}}$ instead of ${\bf{G}}({\bf{x}}){{\bf{G}}^H}({\bf{x}}) \in {{\mathbb{C}}^{N \times N}}$ mainly due to that $K$ is usually smaller than $N$. In other words, a smaller matrix dimension will be more advantageous for the subsequent analysis. Then, based on (46), we can re-express (P3.1) as
 \begin{align}
&({\rm{P3.2}}):{\rm{  }}\mathop {\max }\limits_{{{\bf{x}},{\bf{y}}}} \ {{\bf{y}}^H}{{\bf{G}}^H}({\bf{x}}){\bf{G}}({\bf{x}}){\bf{y}} \tag{${\rm{47a}}$}\\
{\rm{              }}&\ {\rm{s.t.}} \ \quad \ (6{\rm{c}}), (6{\rm{d}}),\tag{${\rm{47b}}$}\\
& \ \ \ \quad \quad \ {{{\left\| {\bf{y}} \right\|}^2} = 1}.\tag{${\rm{47c}}$}
 \end{align}

In the next, AO will be exploited to solve (P3.2) in an alternating manner.

  \begin{figure*}[b!]
  \hrulefill
\setcounter{mytempeqncnt}{\value{equation}}
\setcounter{equation}{58}
\begin{equation}
\begin{split}{}
{{\bf{y}}^H}{{\bf{G}}^H}({\bf{x}}){\bf{G}}({\bf{x}}){\bf{y}} =& {{\bf{y}}^H}\left[ {\begin{array}{*{20}{c}}
{\sqrt {{{\rm{D}}_{1,1}}} {\bf{g}}_1^H({\bf{x}})}\\
 \vdots \\
{\sqrt {{{\rm{D}}_{1,K}}} {\bf{g}}_K^H({\bf{x}})}
\end{array}} \right]\left[ {\begin{array}{*{20}{c}}
{\sqrt {{{\rm{D}}_{1,1}}} {{\bf{g}}_1}({\bf{x}})}&{...}&{\sqrt {{{\rm{D}}_{1,K}}} {{\bf{g}}_K}({\bf{x}})}
\end{array}} \right]{\bf{y}}\\
 = &{\Theta _{10}}({\bf{x}}) = \sum\nolimits_{i = 1}^K {\sum\nolimits_{j = 1}^K {\sqrt {{{\rm{D}}_{1,j}}{{\rm{D}}_{1,i}}} {\mathop{\rm Re}\nolimits} \left( {y_j^H{y_i}{\bf{g}}_j^H({\bf{x}}){{\bf{g}}_i}({\bf{x}})} \right)} } \\
 = &\sum\nolimits_{i = 1}^K {\sum\nolimits_{j = 1}^K {\sum\nolimits_{m = 1}^N {\sqrt {\frac{{{{\rm{D}}_{1,j}}{{\rm{D}}_{1,i}}}}{{d_{{J_j}}^\tau d_{{J_i}}^\tau }}} \left| {y_j^H{y_i}} \right|\cos \left( {\frac{{2\pi }}{\lambda }{x_m}(\sin {\varphi _i} - \sin {\varphi _j}) + {\varphi _{y_j^H{y_i}}}} \right)} } }.
\end{split}
\end{equation}
\setcounter{equation}{\value{mytempeqncnt}}
\vspace{-12pt}
\end{figure*}

\subsection{Optimizing ${\bf{y}}$ Given Fixed ${\bf{x}}$}
When ${\bf{x}}$ is fixed, (P3.2) is simplified as
 \begin{align}
&({\rm{P3.3}}):{\rm{  }}\mathop {\max }\limits_{{{\bf{y}}}} \ {{\bf{y}}^H}{{\bf{G}}^H}({\bf{x}}){\bf{G}}({\bf{x}}){\bf{y}} \tag{${\rm{48a}}$}\\
{\rm{              }}&\ {\rm{s.t.}} \ \quad \ (47{\rm{c}}).\tag{${\rm{48b}}$}
 \end{align}

Since the objective of (P3.3) is non-convex w.r.t. ${\bf{y}}$, the SCA should be exploited to iteratively update this variable until the objective converging to a stationary value. Specifically, given the known ${{\bf{y}}^{(k)}}$ in the $k$-th iteration, ${{\bf{y}}^H}{{\bf{G}}^H}({\bf{x}}){\bf{G}}({\bf{x}}){\bf{y}}$ has the lower bound as
\begin{equation}
\setcounter{equation}{49}
\begin{split}{}
{{\bf{y}}^H}{{\bf{G}}^H}({\bf{x}}){\bf{G}}({\bf{x}}){\bf{y}} \ge & 2{\rm{Re}}\left( {{{\bf{y}}^H}\underbrace {{{\bf{G}}^H}({\bf{x}}){\bf{G}}({\bf{x}}){{\bf{y}}^{(k)}}}_{{\bf{b}}({{\bf{y}}^{(k)}})}} \right)\\
 &- {({{\bf{y}}^{(k)}})^H}{{\bf{G}}^H}({\bf{x}}){\bf{G}}({\bf{x}}){{\bf{y}}^{(k)}},
 \end{split}
\end{equation}
where ${\mathop{\rm Re}\nolimits} \left( {{{\bf{y}}^H}{\bf{b}}({{\bf{y}}^{(k)}})} \right)$ can be expanded as
\begin{equation}
\begin{split}{}
&{\mathop{\rm Re}\nolimits} \left( {{{\bf{y}}^H}{\bf{b}}({{\bf{y}}^{(k)}})} \right) = {\mathop{\rm Re}\nolimits} \left( {\sum\nolimits_{i = 1}^K {y_i^H{b_i}({{\bf{y}}^{(k)}})} } \right)\\
 =& {\mathop{\rm Re}\nolimits} \left( {\sum\nolimits_{i = 1}^K {\left| {y_i^H} \right|\left| {{b_i}({{\bf{y}}^{(k)}})} \right|{e^{j\left( {{\varphi _{y_i^H}} + {\varphi _{{b_i}({{\bf{y}}^{(k)}})}}} \right)}}} } \right)\\
  =& {\Theta _9}({\bf{y}},{{\bf{y}}^{(k)}}),
 \end{split}
\end{equation}
with ${y_i^H}$ and ${{b_i}({{\bf{y}}^{(k)}})}$ denoting the $i$-th element in ${{{\bf{y}}^H}}$ and the $i$-th element in ${{\bf{b}}({{\bf{y}}^{(k)}})}$, and ${{\varphi _{y_i^H}}}$ and ${{\varphi _{{b_i}({{\bf{y}}^{(k)}})}}}$ denoting the phases of ${y_i^H}$ and ${{b_i}({{\bf{y}}^{(k)}})}$, respectively. Further, note that
\begin{equation}
{\left\| {\bf{y}} \right\|^2} = 1 \Leftrightarrow \sum\nolimits_{i = 1}^K {{{\left| {y_i^H} \right|}^2}}  = 1 \ {\rm{and}} \ \left\{ {{\varphi _{y_i^H}}} \right\}_{i = 1}^K \in [0,2\pi ].
 \end{equation}

Based on (49)$-$(51) and ignoring the irrelevant constant, in the $k$-th iteration, we just need to solve the following problem
 \begin{align}
&({\rm{P3.4}}):{\rm{  }}\mathop {\max }\limits_{{{\bf{y}}}} \ {\Theta _9}({\bf{y}},{{\bf{y}}^{(k)}}) \tag{${\rm{52a}}$}\\
{\rm{              }}&\ {\rm{s.t.}} \ \quad \ (51).\tag{${\rm{52b}}$}
 \end{align}

To solve (P3.4), from (50) we can first judge that to maximize ${\Theta _9}({\bf{y}},{{\bf{y}}^{(k)}})$, ${\varphi _{y_i^H}}$, $i = 1,...,K$, should be set as
\begin{equation}
\setcounter{equation}{53}
{\varphi _{y_i^H}} =  - {\varphi _{{b_i}({{\bf{y}}^{(k)}})}} + l,
 \end{equation}
with $l$ denoting an arbitrary constant. Based on (53), ${\Theta _9}({\bf{y}},{{\bf{y}}^{(k)}})$ can be simplified as ${\Theta _9}({\bf{y}},{{\bf{y}}^{(k)}}) = \sum\nolimits_{i = 1}^K {\left| {y_i^H} \right|\left| {{b_i}({{\bf{y}}^{(k)}})} \right|} $. Then, (P3.4) can be further simplified as
 \begin{align}
&({\rm{P3.5}}):{\rm{  }}\mathop {\max }\limits_{\left\{ {\left| {y_i^H} \right|} \right\}} \ \sum\nolimits_{i = 1}^K {\left| {y_i^H} \right|\left| {{b_i}({{\bf{y}}^{(k)}})} \right|}  \tag{${\rm{54a}}$}\\
{\rm{              }}&\ {\rm{s.t.}} \ \quad \ \sum\nolimits_{i = 1}^K {{{\left| {y_i^H} \right|}^2}}  = 1.\tag{${\rm{54b}}$}
 \end{align}

Note that (P3.5) is a convex problem and its corresponding Lagrange function is given by
\begin{equation}
\setcounter{equation}{55}
 \begin{split}
{\cal L}(\left\{ {\left| {y_i^H} \right|} \right\},\zeta ) =& \sum\nolimits_{i = 1}^K {\left| {y_i^H} \right|\left| {{b_i}({{\bf{y}}^{(k)}})} \right|} \\
 &- \zeta \left( {\sum\nolimits_{i = 1}^K {{{\left| {y_i^H} \right|}^2}}  - 1} \right),
 \end{split}
 \end{equation}
where $\zeta  \ge 0$ is the Lagrange multiplier associated with the constraint in (54b). Then, by solving the following Lagrange equations
\begin{equation}
\left\{ {\begin{array}{*{20}{c}}
{\frac{{\partial {\cal L}(\left\{ {\left| {y_i^H} \right|} \right\},\zeta )}}{{\partial \left| {y_i^H} \right|}} = \left| {{b_i}({{\bf{y}}^{(k)}})} \right| - 2\zeta \left| {y_i^H} \right| = 0,i = 1,...,K}\\
{\frac{{\partial {\cal L}(\left\{ {\left| {y_i^H} \right|} \right\},\zeta )}}{{\partial \zeta }} = \sum\nolimits_{i = 1}^K {{{\left| {y_i^H} \right|}^2}}  - 1 = 0}
\end{array}} \right.,
 \end{equation}
 we can obtain the optimal $\left| {y_i^H} \right|$ (denoted as ${\left| {y_i^H} \right|^{({\rm{opt}})}}$) as
 \begin{equation}
 {\left| {y_i^H} \right|^{({\rm{opt}})}} = \frac{{\left| {{b_i}({{\bf{y}}^{(k)}})} \right|}}{{\sqrt {\sum\nolimits_{j = 1}^K {{{\left| {{b_j}({{\bf{y}}^{(k)}})} \right|}^2}} } }},i = 1,...,K.
  \end{equation}

Based on (53) and (57), we now can finally determine that given ${{{\bf{y}}^{(k)}}}$, the optimal ${\bf{y}}$ (denoted as ${{\bf{y}}^{({\rm{opt}})}}$) to (P3.4) can be expressed as
 \begin{equation}
{{\bf{y}}^{({\rm{opt}})}} = {\left[ {{{\left| {y_1^H} \right|}^{({\rm{opt}})}}{e^{j({\varphi _{{b_1}({{\bf{y}}^{(k)}})}} - l)}},...,{{\left| {y_K^H} \right|}^{({\rm{opt}})}}{e^{j({\varphi _{{b_K}({{\bf{y}}^{(k)}})}} - l)}}} \right]^T}.
  \end{equation}

Complexity analysis: In each iteration, since the optimal ${\bf{y}}$ has the closed-form solution, the complexity of solving (P3.3) is only ${\cal O}({I_{{\rm{iter}},1}})$, with ${I_{{\rm{iter}},1}}$ denoting the number of iterations.

\begin{algorithm}
\caption{The AO for Solving (P3.2)}
  \begin{algorithmic}[1]

\State \textbf{{Input:}} ${{\bf{x}}^{(1)}}$, ${{\bf{y}}^{(1)}}$ and $k = 1$.

\State \textbf{Repeat:}


\State \quad \textbf{Repeat:}

\State \quad Solve (P3.5) and output ${{\bf{y}}^{({\rm{opt}})}}$. 

\State \quad $k \leftarrow k + 1$.

\State \quad Set ${{\bf{y}}^{(k)}} = {{\bf{y}}^{({\rm{opt}})}}$.

\State \quad \textbf{Until:} ${\Theta _9}({\bf{y}},{{\bf{y}}^{(k)}})$ converges to a stationary value.


\State \quad \textbf{Repeat:}

\State \quad Solve (P3.7) and output ${{\bf{x}}}$. 

\State \quad $k \leftarrow k + 1$.

\State \quad Set ${\bf{x}}^{(k)} = {{\bf{x}}}$.

\State \quad \textbf{Until:} ${\Theta _{10}}({\bf{x}},{{\bf{x}}^{(k)}})$ converges to a stationary value.


\State \textbf{Until:} $\left| {{\Theta _{10}}({\bf{x}},{{\bf{x}}^{(k - 1)}}) - {\Theta _9}({\bf{y}},{{\bf{y}}^{(k - 1)}})} \right| \le \varepsilon $.

\State \textbf{Output:} ${\bf{x}}$ and ${\bf{y}}$.
  \end{algorithmic}
\end{algorithm}

\subsection{Optimizing ${\bf{x}}$ Given Fixed ${\bf{y}}$}
Given ${\bf{y}}$, we first expand the objective of (P3.2) as in (59), where ${{\varphi _{y_j^H{y_i}}}}$ is the phase of the complex scalar ${y_j^H{y_i}}$. Based on (59), (P3.2) can be simplified as
 \begin{align}
&({\rm{P3.6}}):{\rm{  }}\mathop {\max }\limits_{{{\bf{x}}}} \ {\Theta _{10}}({\bf{x}}) \tag{${\rm{60a}}$}\\
{\rm{              }}&\ {\rm{s.t.}} \ \quad \ (47{\rm{b}}).\tag{${\rm{60b}}$}
 \end{align}

Then, applying the similar method in the above, given the known ${{\bf{x}}^{(k)}}$, the lower bound of ${\Theta _{10}}({\bf{x}})$ can be derived as
\begin{equation}
\setcounter{equation}{61}
 \begin{split}
&{\Theta _{10}}({\bf{x}}) \ge {\Theta _{10}}({\bf{x}},{{\bf{x}}^{(k)}})\\
 =& \sum\nolimits_{i = 1}^K {\sum\nolimits_{j = 1}^K {\sum\nolimits_{m = 1}^N {\sqrt {\frac{{{{\rm{D}}_{1,j}}{{\rm{D}}_{1,i}}}}{{d_{{J_j}}^\tau d_{{J_i}}^\tau }}} \left| {y_j^H{y_i}} \right|{\Theta _{{11},i,j}}({x_m},x_m^{(k)})} } },
 \end{split}
 \end{equation}
with
\begin{equation}
 \begin{split}
&{\Theta _{{11},i,j}}({x_m},x_m^{(k)})\\
 =& \cos \left( {\frac{{2\pi }}{\lambda }x_m^{(k)}(\sin {\varphi _i} - \sin {\varphi _j}) + {\varphi _{y_j^H{y_i}}}} \right)\\
 &- \frac{{2\pi }}{\lambda }(\sin {\varphi _i} - \sin {\varphi _j})\\
 &\times \sin \left( {\frac{{2\pi }}{\lambda }x_m^{(k)}(\sin {\varphi _i} - \sin {\varphi _j}) + {\varphi _{y_j^H{y_i}}}} \right)\\
 &\times ({x_m} - x_m^{(k)}) - \frac{{{\delta _{i,j}}}}{2}{\left( {{x_m} - x_m^{(k)}} \right)^2},
 \end{split}
 \end{equation}
where ${\delta _{i,j}} = {\left( {\frac{{2\pi }}{\lambda }(\sin {\varphi _i} - \sin {\varphi _j})} \right)^2}$.

Armed with the above analysis, (P3.6) can be transformed to the following approximate problem
  \begin{align}
&({\rm{P3.7}}):{\rm{  }}\mathop {\max }\limits_{{{\bf{x}}}} \ {\Theta _{10}}({\bf{x}},{{\bf{x}}^{(k)}}) \tag{${\rm{63a}}$}\\
{\rm{              }}&\ {\rm{s.t.}} \ \quad \ (47{\rm{b}}).\tag{${\rm{63b}}$}
 \end{align}
Note that (P3.7) is a convex optimization problem, which can be iteratively solved by using CVX.

Complexity analysis: Since there are $N$ real variables in (P3.7), the corresponding complexity is about ${\cal O}({I_{{\rm{iter}},2}}{N^{3.5}})$, where ${{I_{{\rm{iter}},2}}}$ is the number of iterations.

\subsection{Overall Algorithm, Complexity and Convergence}
We now summarize our proposed AO algorithm for solving (P3.2), as shown in Algorithm 2. The initial input includes the known ${{\bf{x}}^{(1)}}$ and ${{\bf{y}}^{(1)}}$. Afterwards, in each outer iteration, (P3.5) and (P3.7) are solved sequentially by implementing steps 3$-$7 and steps 8$-$12 in Algorithm 2, respectively. Finally, the outer iteration can be terminated if the condition in step 13 is satisfied. In addition, the total complexity of Algorithm 2 is about ${\cal O}({I_{{\rm{iter}},{\rm{out}}}}\left( {{I_{{\rm{iter}},1}} + {I_{{\rm{iter}},2}}{N^{3.5}}} \right))$, where ${I_{{\rm{iter}},{\rm{out}}}}$ is the number of outer iterations. The algorithm convergence can be analyzed using the similar method in Section III, thus the details are not repeated here.

\subsection{Initialization}
In this subsection, we further provide the detailed initialization scheme for Algorithm 2. Specifically, during the initialization phase: i) the positions of MAs at the MAJ are also set as ${{\bf{x}}^{(1)}} = \left[ {0:L/(N - 1):L} \right]$. The reasons are presented in the above and are omitted here; ii) the slack variables are set as ${{\bf{y}}^{(1)}} = \left( {{\bf{s}} + j{\bf{s}}} \right)/\left\| {{\bf{s}} + j {\bf{s}}} \right\|$ through comparative experiments, where ${\bf{s}} = {\left[ {1,...,1} \right]^T} \in {{\mathbb{R}}^{K \times 1}}$.

\section{Insights from the Special Case of $K = 2$}
In this section, focusing on the special case of $K = 2$, i.e., there are two suspicious receivers ${\rm{R}}_1$ and ${\rm{R}}_2$, we aim to reveal some insightful conclusions for better understanding the deployment rule of antenna positions at the MAJ.

\subsection{Minimizing the Sum Rate}
In this scenario, even when $K = 2$, we still cannot determine the optimal forms of ${{\bf{w}}_J}$ and ${\bf{x}}$ from (P2) immediately, as the objective and constraints of (P2) have complex structures. Therefore, we will approach the solution from another perspective as follows.

Specifically, when $K = 2$, for the MAJ, it should consider how to achieve the optimal jamming balance to these two suspicious receivers, by jointly optimizing ${{\bf{w}}_J}$ and ${\bf{x}}$. From this observation, the MAJ should care about the following problem
 \begin{align}
&({\rm{P4}}):{\rm{  }}\mathop {\max }\limits_{{{\bf{w}}_J},{\bf{x}}} \ a{\left| {{\bf{w}}_J^H{{\bf{g}}_1}({\bf{x}})} \right|^2} + (1 - a){\left| {{\bf{w}}_J^H{{\bf{g}}_2}({\bf{x}})} \right|^2} \tag{${\rm{64a}}$}\\
{\rm{              }}&\ {\rm{s.t.}} \quad (6{\rm{b}}), (6{\rm{c}}), (6{\rm{d}}),\tag{${\rm{64b}}$}
 \end{align}
where $a \in [0,1]$ is the jamming weight coefficient to be optimized. For instance, when $a = 1$, the MAJ cares about maximizing the jamming effect to ${\rm{R}}_1$; otherwise when $a = 0$, the MAJ will maximize the jamming effect to ${\rm{R}}_2$. By adaptively adjusting $a$, the MAJ can achieve different jamming balance patterns. Note that $a$ should be optimized by further considering the flexible power allocations of the ST, the details will be explained later.

  \begin{figure*}[b!]
  \hrulefill
\setcounter{mytempeqncnt}{\value{equation}}
\setcounter{equation}{73}
\begin{equation}
\begin{split}{}
\left| {\xi {{\bf{I}}_K} - {\bf{G}}_K^H({\bf{x}}){{\bf{G}}_K}({\bf{x}})} \right| = 0 \Leftrightarrow \left| {\begin{array}{*{20}{c}}
{\xi  - {a_1}N/d_{{J_1}}^\tau }&{ - \sqrt {{a_1}{a_2}} {r_{1,2}}({\bf{x}})}& \cdots &{ - \sqrt {{a_1}{a_K}} {r_{1,K}}({\bf{x}})}\\
{ - \sqrt {{a_1}{a_2}} {r_{2,1}}({\bf{x}})}&{\xi  - {a_2}N/d_{{J_2}}^\tau }& \cdots &{ - \sqrt {{a_2}{a_K}} {r_{2,K}}({\bf{x}})}\\
 \vdots & \vdots & \ddots & \vdots \\
{ - \sqrt {{a_1}{a_K}} {r_{K,1}}({\bf{x}})}& \cdots & \cdots &{\xi  - {a_K}N/d_{{J_K}}^\tau }
\end{array}} \right| = 0.
\end{split}
\end{equation}
\setcounter{equation}{\value{mytempeqncnt}}
\vspace{-12pt}
\end{figure*}

Now, by fixing $a$, we focus on how to solve (P4). To proceed, using the same method presented in Section IV, we directly re-express the objective of (P4) as
\begin{equation}
\setcounter{equation}{65}
\begin{split}{}
&a{\left| {{\bf{w}}_J^H{{\bf{g}}_1}({\bf{x}})} \right|^2} + (1 - a){\left| {{\bf{w}}_J^H{{\bf{g}}_2}({\bf{x}})} \right|^2}\\
 =& {\rm{Tr}}\left( {{{\bf{G}}_2 }({\bf{x}}){\bf{G}}_2 ^H({\bf{x}}){{\bf{W}}_J}} \right),
 \end{split}
\end{equation}
where ${{{\bf{G}}_2 }({\bf{x}})} = \left[ {\sqrt a {{\bf{g}}_1}({\bf{x}}),\sqrt {1 - a} {{\bf{g}}_2}({\bf{x}})} \right] \in {{\mathbb{C}}^{N \times 2}}$. Similarly, by deriving the optimal ${{\bf{W}}_J}$ based on (43) and then substituting it into (65), the objective of (P4) is simplified as
\begin{equation}
\begin{split}{}
{\rm{Tr}}\left( {{{\bf{G}}_2 }({\bf{x}}){\bf{G}}_2 ^H({\bf{x}}){\bf{W}}_J^*} \right) =& {\xi _{\max }}\left( {{{\bf{G}}_2 }({\bf{x}}){\bf{G}}_2 ^H({\bf{x}})} \right)\\
 =& {\xi _{\max }}\left( {{\bf{G}}_2 ^H({\bf{x}}){{\bf{G}}_2 }({\bf{x}})} \right),
 \end{split}
\end{equation}
where ${{\bf{G}}_2 ^H({\bf{x}}){{\bf{G}}_2 }({\bf{x}})} \in {{\mathbb{C}}^{2 \times 2}}$ based on the expression of ${{{\bf{G}}_2 }({\bf{x}})}$ can be expressed as
\begin{equation}
\begin{split}{}
&{{\bf{G}}_2 ^H({\bf{x}}){{\bf{G}}_2 }({\bf{x}})}\\
 =& \left[ {\begin{array}{*{20}{c}}
{aN/d_{{J_1}}^\tau }&{\sqrt {a(1 - a)} r({\bf{x}})}\\
{\sqrt {a(1 - a)} {r^*}({\bf{x}})}&{(1 - a)N/d_{{J_2}}^\tau }
\end{array}} \right],
\end{split}
\end{equation}
where $r({\bf{x}}) \buildrel \Delta \over = {\bf{g}}_1^H({\bf{x}}){{\bf{g}}_2}({\bf{x}})$.

Based on (67), we can determine that the principal eigenvalue of ${{\bf{G}}_2 ^H({\bf{x}}){{\bf{G}}_2 }({\bf{x}})}$ is the maximum root of the following equation
\begin{equation}
\left| {\xi {{\bf{I}}_2} - {\bf{G}}_2 ^H({\bf{x}}){{\bf{G}}_2 }({\bf{x}})} \right| = 0,
\end{equation}
which is equivalent to
\begin{equation}
\begin{split}{}
&\left| {\begin{array}{*{20}{c}}
{\xi  - aN/d_{{J_1}}^\tau }&{ - \sqrt {a(1 - a)} r({\bf{x}})}\\
{ - \sqrt {a(1 - a)} {r^*}({\bf{x}})}&{\xi  - (1 - a)N/d_{{J_2}}^\tau }
\end{array}} \right| = 0,\\
 \Leftrightarrow &(\xi  - aN/d_{{J_1}}^\tau )(\xi  - (1 - a)N/d_{{J_2}}^\tau )\\
& \quad \quad \quad \quad \quad \quad \quad \quad \quad \quad \ \ - a(1 - a){\left| {r({\bf{x}})} \right|^2} = 0.
\end{split}
\end{equation}


 \begin{figure}[!t]
\vspace{-15pt}
\centering
\includegraphics[width=7cm]{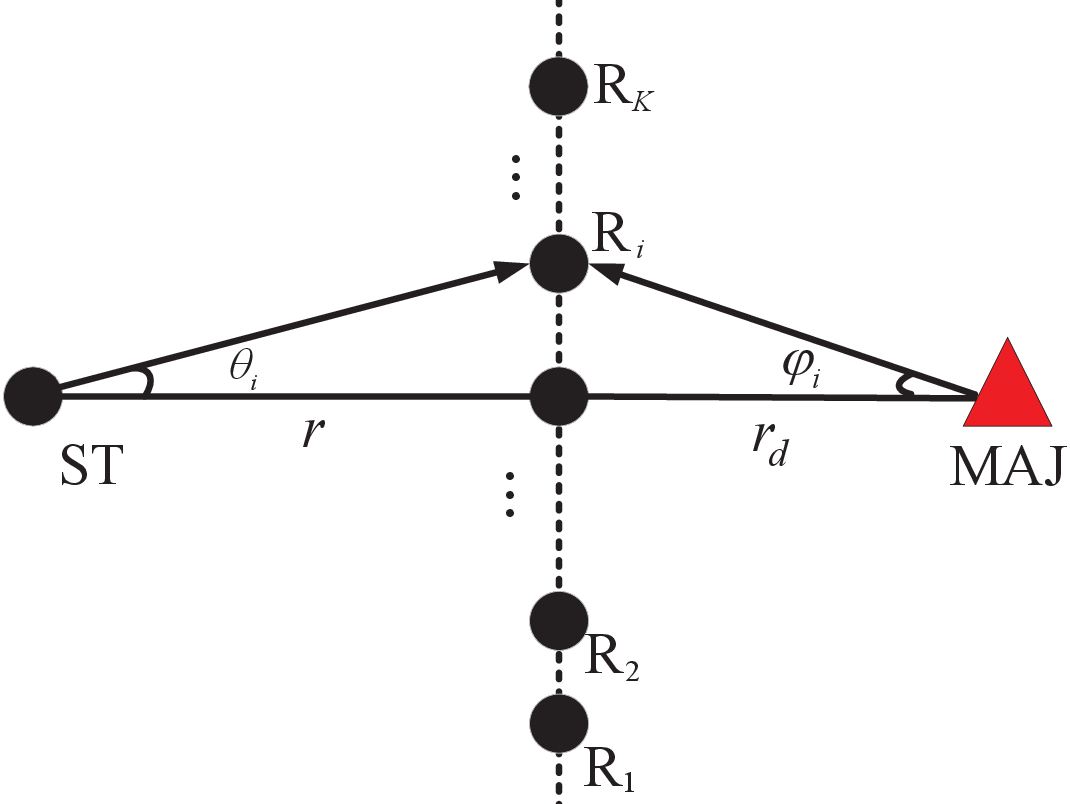}
\captionsetup{font=small}
\caption{Simulation setups.}
\label{fig:Fig1}
\end{figure}

By solving the equation in (69), we can compute that
\begin{equation}
\begin{split}{}
&{\xi _{\max }}\left( {{\bf{G}}_2 ^H({\bf{x}}){{\bf{G}}_2 }({\bf{x}})} \right)\\
 =& \frac{{f(a) + \sqrt {{f^2}(a) - 4a(1 - a)\left( {\frac{{{N^2}}}{{d_{{J_1}}^\tau d_{{J_2}}^\tau }} - {{\left| {r({\bf{x}})} \right|}^2}} \right)} }}{2},
\end{split}
\end{equation}
where $f(a) = aN/d_{{J_1}}^\tau  + (1 - a)N/d_{{J_2}}^\tau $.

From (70), we can clearly observe that ${\xi _{\max }}\left( {{\bf{G}}_2 ^H({\bf{x}}){{\bf{G}}_2 }({\bf{x}})} \right)$ is monotonically increasing w.r.t. ${\left| {r({\bf{x}})} \right|^2} \buildrel \Delta \over = {\left| {{\bf{g}}_1^H({\bf{x}}){{\bf{g}}_2}({\bf{x}})} \right|^2}$. Based on this key finding, (P4) can be further simplified as
 \begin{align}
&({\rm{P4.1}}):{\rm{  }}\mathop {\max }\limits_{{\bf{x}}} \ {\left| {{\bf{g}}_1^H({\bf{x}}){{\bf{g}}_2}({\bf{x}})} \right|^2}\tag{${\rm{71a}}$}\\
{\rm{              }}&\ {\rm{s.t.}} \ \quad \ (6{\rm{c}}), (6{\rm{d}}).\tag{${\rm{71b}}$}
 \end{align}

From (P4.1), it is interesting to see that in the special case of $K = 2$, for any given weight $a$, antenna positions at the MAJ should be definitely optimized based on the rule of maximizing the correlation between the two jamming channels ${{{\bf{g}}_1}({\bf{x}})}$ and ${{{\bf{g}}_2}({\bf{x}})}$. This finding actually is intuitive, since such rule creates the best prerequisite for the MAJ to optimize its jamming beamforming for achieving the best jamming balance.

Note that (P4.1) can be handled by deriving the lower bound of its objective and then exploiting the SCA. The details have been introduced in the above and thus are not repeated here for brevity.

Based on the above analysis, we summarize the optimization process for the special case of $K = 2$ in the follows.
\begin{itemize}
\item[$\bullet$]  First, antenna positions at the MAJ are optimized by solving (P4.1). Denote the output as ${{\bf{x}}^*}$.

 \item[$\bullet$] Second, set $a = 0:\varepsilon :1$ ($\varepsilon$ is the search accuracy). Based on (43), obtain the corresponding optimal jamming beamforming as ${\bf{w}}_J^*(a) = {{\bf{u}}_{\max }}\left( {{{\bf{G}}_2 }({{\bf{x}}^*}){\bf{G}}_2 ^H({{\bf{x}}^*})} \right)$.

 \item[$\bullet$]  Third, given each $a$, substituting ${{\bf{x}}^*}$ and the corresponding ${\bf{w}}_J^*(a)$ into (11) to compute the water-filling level (denoted as $\eta (a)$). Then, substituting ${{\bf{x}}^*}$, ${\bf{w}}_J^*(a)$ and $\eta (a)$ into the objective in (10) to obtain the suspicious sum rate (denoted as ${R_{{\rm{sum}}}}(a)$).

 \item[$\bullet$] Finally, focus on the set ${\left\{ {{R_{{\rm{sum}}}}(a)} \right\}_{a = 0:\varepsilon :1}}$, and select the minimum value and the corresponding $a$ (denoted as ${a^*}$). Output the optimal jamming beamforming as ${\bf{w}}_J^*(a^*)$.
\end{itemize}

Complexity analysis: For the first step in the above, using the SCA, the complexity of obtaining ${{{\bf{x}}^*}}$ can be similarly concluded as ${\cal O}({I_{{\rm{iter}}}}{N^{3.5}})$, with ${I_{{\rm{iter}}}}$ denoting the number of iterations. For the second step, given each $a$, since obtaining ${\bf{w}}_J^*(a)$ requires the eigenvalue decomposition on ${{{\bf{G}}_2 }({{\bf{x}}^*}){\bf{G}}_2 ^H({{\bf{x}}^*})}$, the corresponding complexity is about ${\cal O}({N^2})$. Then, by exhausting all cases of $a = 0:\varepsilon :1$, the complexity is about ${\cal O}({N^2}/\varepsilon )$. Note that step 1 and step 2 are independent of each other. Therefore, the total complexity of obtaining ${{\bf{x}}^*}$ and ${\bf{w}}_J^*({a^*})$ is about ${\cal O}({I_{{\rm{iter}}}}{N^{3.5}} + {N^2}/\varepsilon )$.

\subsection{Minimizing the Fairness}
In this scenario, when $K = 2$, the globally optimal ${{\bf{w}}_J}$ can be immediately obtained from (43) due to its clear structure. Afterwards, based on (45) and the above analysis in subsection A, we can directly verify that antenna positions at the MAJ should be also optimized based on the rule of maximizing the jamming channel correlation. The total complexity of obtaining ${{\bf{x}}^*}$ and ${\bf{w}}_J^*$ in this scenario is only about ${\cal O}({I_{{\rm{iter}}}}{N^{3.5}} + 1/\varepsilon )$.

\subsection{Extension to the General Multiuser Case}
When the number of users is arbitrary $K > 2$, problem (P4) can be reformulated as
\begin{align}
&({\rm{P5}}):{\rm{  }}\mathop {\max }\limits_{{{\bf{w}}_J},{\bf{x}}} \ \sum\nolimits_{i = 1}^K {{a_i}{{\left| {{\bf{w}}_J^H{{\bf{g}}_i}({\bf{x}})} \right|}^2}}  \tag{${\rm{72a}}$}\\
{\rm{              }}&\ {\rm{s.t.}} \quad (6{\rm{b}}), (6{\rm{c}}), (6{\rm{d}}),\tag{${\rm{72b}}$}
 \end{align}

\noindent where $\left\{ {{a_i}} \right\}_{i = 1}^K \in [0,1]$ are the weight coefficients to be optimized, which satisfy $\sum\nolimits_{i = 1}^K {{a_i}}  = 1$. Then, following the same setups presented in subsection A, we can determine that (P5) can be equivalently formulated as
\begin{align}
&({\rm{P5.1}}):{\rm{  }}\mathop {\max }\limits_{{\bf{x}}} \ {\xi _{\max }}\left( {{\bf{G}}_K^H({\bf{x}}){{\bf{G}}_K}({\bf{x}})} \right)  \tag{${\rm{73a}}$}\\
{\rm{              }}&\ {\rm{s.t.}} \quad (6{\rm{b}}), (6{\rm{c}}), (6{\rm{d}}),\tag{${\rm{73b}}$}
 \end{align}

 \noindent where ${{\bf{G}}_K}({\bf{x}}) = \left[ {\sqrt {{a_1}} {{\bf{g}}_1}({\bf{x}}),\sqrt {{a_2}} {{\bf{g}}_2}({\bf{x}}),...,\sqrt {{a_K}} {{\bf{g}}_K}({\bf{x}})} \right] \in {{\mathbb{C}}^{N \times K}}$, and ${\xi _{\max }}\left( {{\bf{G}}_K^H({\bf{x}}){{\bf{G}}_K}({\bf{x}})} \right)$ is the maximum root of equation (74), with ${r_{i,j}}({\bf{x}}) = \left| {{\bf{g}}_i^H({\bf{x}}){{\bf{g}}_j}({\bf{x}})} \right|$, $i,j = 1,...,K$ and $i \ne j$. Unfortunately, unlike the case of $K = 2$, the complex structure shown in (74) hinders the derivation for the closed-form ${\xi _{\max }}\left( {{\bf{G}}_K^H({\bf{x}}){{\bf{G}}_K}({\bf{x}})} \right)$. However, from (74), we can judge that given $\left\{ {{a_i}} \right\}_{i = 1}^K$, ${\xi _{\max }}\left( {{\bf{G}}_K^H({\bf{x}}){{\bf{G}}_K}({\bf{x}})} \right)$ should be the function of $\left\{ {{r_{i,j}}({\bf{x}})} \right\}_{i,j = 1,i \ne j}^K$, i.e., ${\xi _{\max }}\left( {{\bf{G}}_K^H({\bf{x}}){{\bf{G}}_K}({\bf{x}})} \right) = f\left( {\left\{ {{r_{i,j}}({\bf{x}})} \right\}_{i,j = 1,i \ne j}^K} \right)$. Based on this fact, we can also design the sub-optimal antenna positions from the perspective of varying numerous rules related to channel correlations, such as $\mathop {\max }\limits_{\bf{x}} \mathop {\min }\limits_{i,j = 1,...,k,i \ne j} {r_{i,j}}({\bf{x}})$ and $\mathop {\max }\limits_{\bf{x}} \sum\nolimits_{i = 1}^K {\sum\nolimits_{j = 1,j \ne i}^K {{r_{i,j}}({\bf{x}})} } $.

\begin{figure}
\begin{center}

\begin{minipage}{7.9cm}
\includegraphics[width=7.9cm]{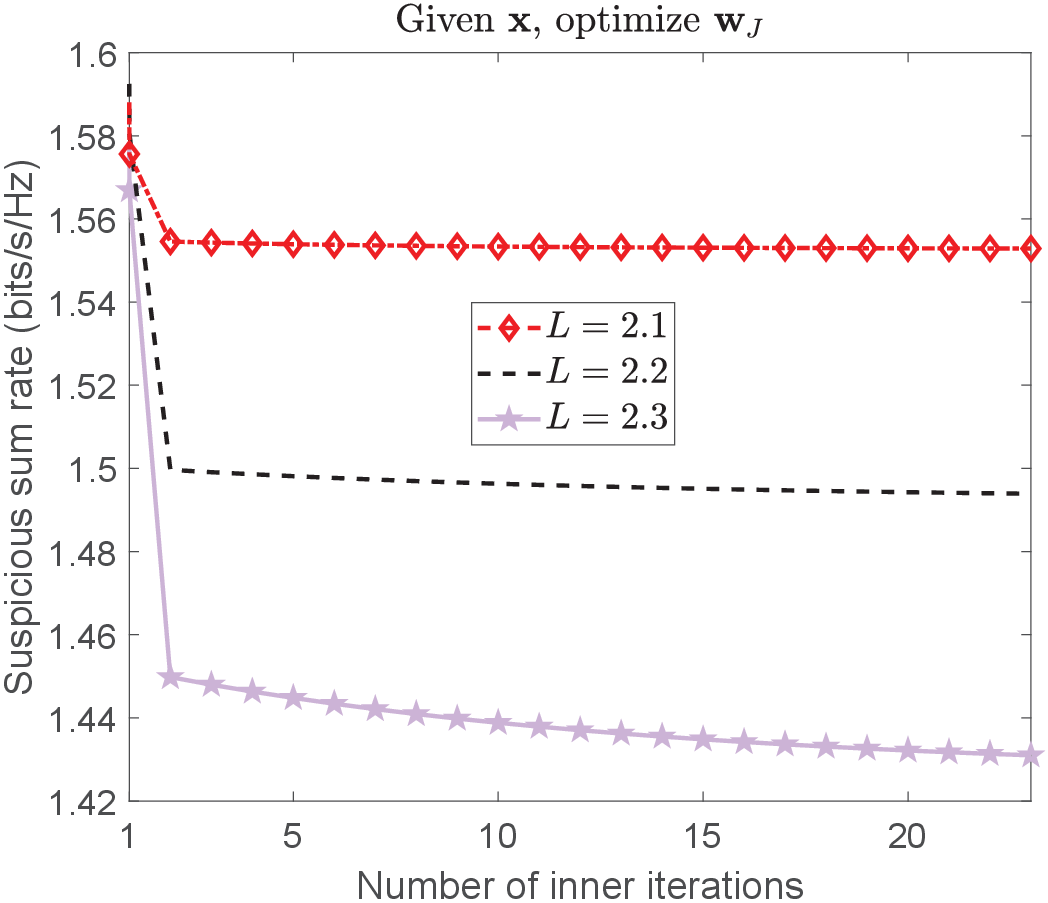}
\centering
\subfigure{(a)}

\end{minipage}
\begin{minipage}{7.9cm}
\includegraphics[width=7.9cm]{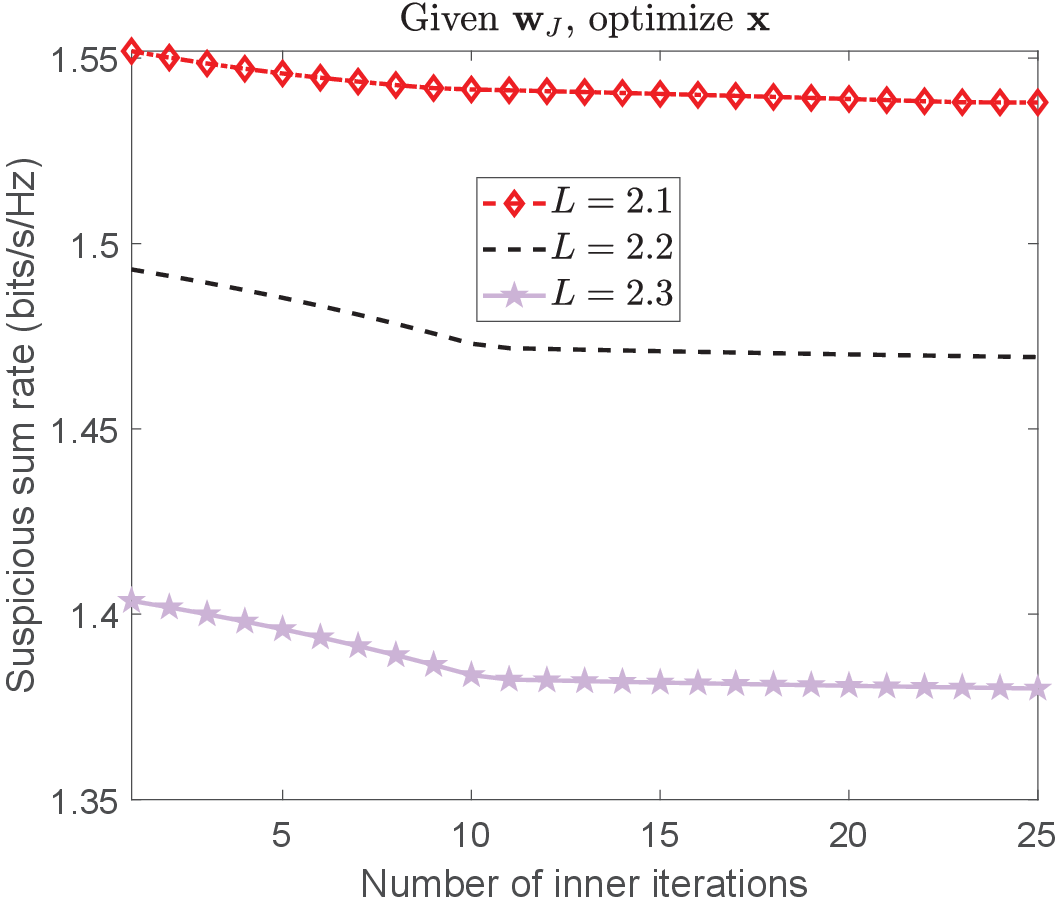}
\centering
\subfigure{(b)}

\end{minipage}
\begin{minipage}{7.9cm}
\includegraphics[width=7.9cm]{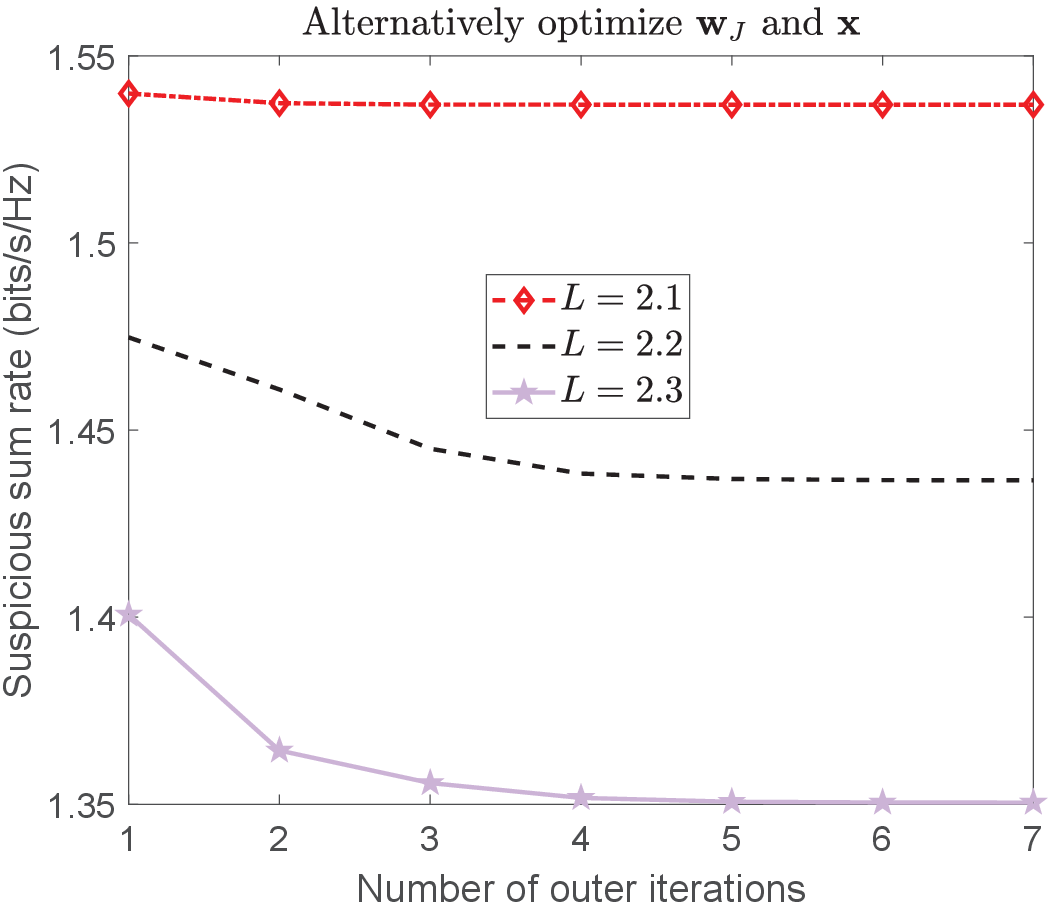}
\centering
\subfigure{(c)}

\end{minipage}
\captionsetup{font=small}
\caption{Convergence behaviors of Algorithm 1. The inner iteration: (a) given ${\bf{x}}$ and optimize ${{\bf{w}}_J}$; (b) given ${{\bf{w}}_J}$ and optimize ${\bf{x}}$. The outer iteration: (c) alternatively optimize ${{\bf{w}}_J}$ and ${\bf{x}}$.}
\end{center}
\end{figure}

\begin{figure}
\begin{center}

\begin{minipage}{7.9cm}
\includegraphics[width=7.9cm]{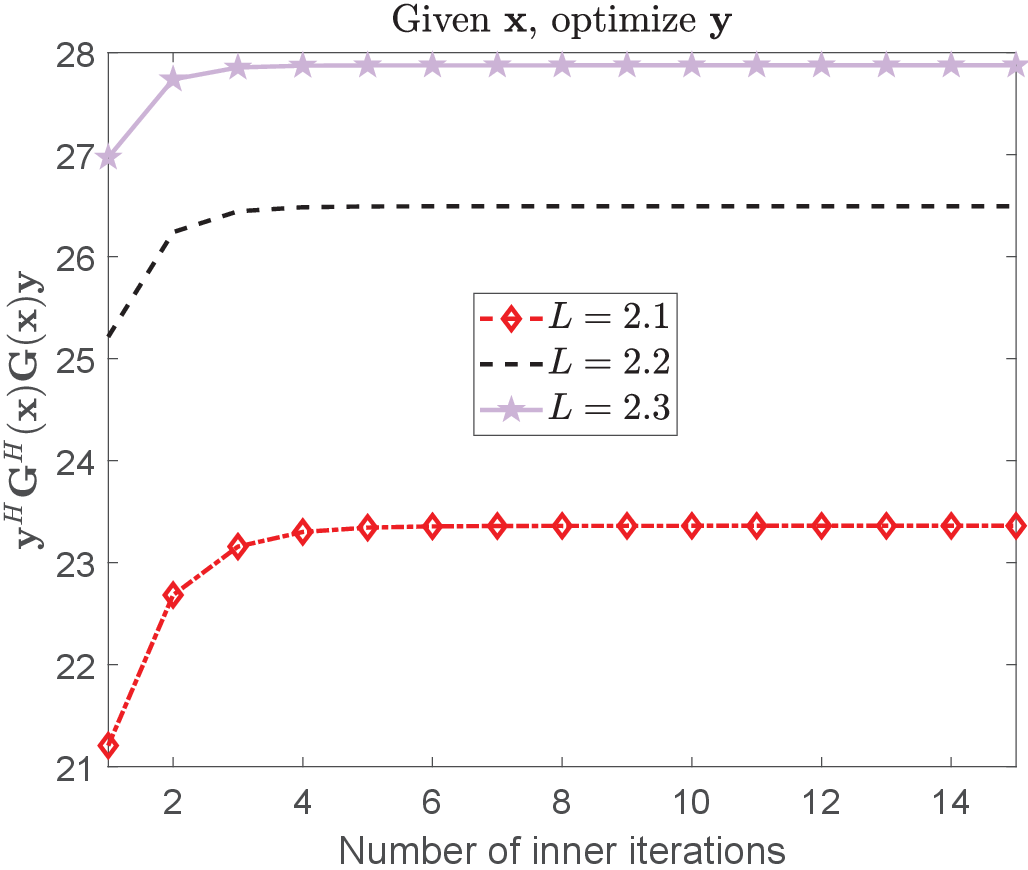}
\centering
\subfigure{(a)}

\end{minipage}
\begin{minipage}{7.9cm}
\includegraphics[width=7.9cm]{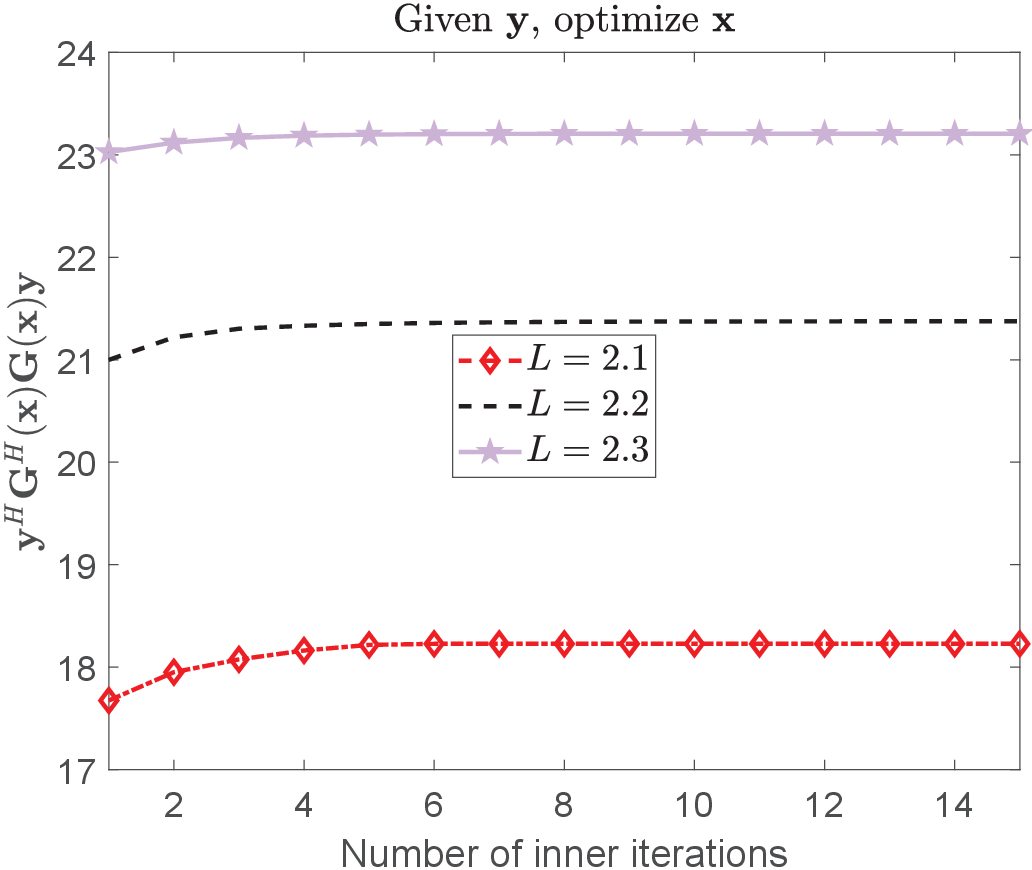}
\centering
\subfigure{(b)}

\end{minipage}
\begin{minipage}{7.9cm}
\includegraphics[width=7.9cm]{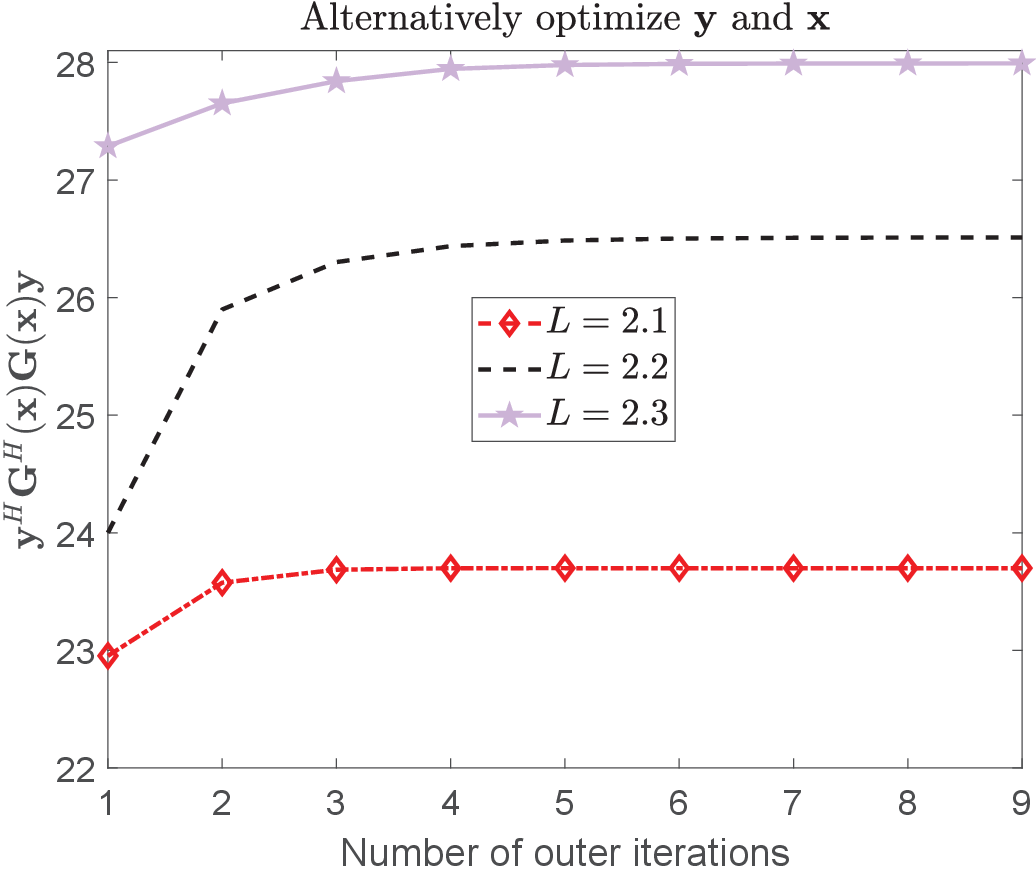}
\centering
\subfigure{(c)}

\end{minipage}
\captionsetup{font=small}
\caption{Convergence behaviors of Algorithm 2. The inner iteration: (a) given ${\bf{x}}$ and optimize ${\bf{y}}$; (b) given ${\bf{y}}$ and optimize ${\bf{x}}$. The outer iteration: (c) alternatively optimize ${\bf{y}}$ and ${\bf{x}}$.}
\end{center}
\end{figure}

\section{Performance Bounds}
From the perspective of the MAJ, when the correlation between any two jamming channels achieves the global maximum, i.e., $\left| {{\bf{g}}_i^H({\bf{x}}){{\bf{g}}_j}({\bf{x}})} \right|/N = 1/\sqrt {d_{{J_i}}^\tau d_{{J_j}}^\tau } $, $\forall i,j = 1,...,K,i \ne j$, the jamming effect to all suspicious receivers can be maximized by simply setting the jamming beamforming as the form of maximum-ratio transmission (MRT), i.e., ${\bf{w}}_J^{{\rm{MRT}}} = {{\bf{g}}_i}({\bf{x}})/\left\| {{{\bf{g}}_i}({\bf{x}})} \right\|$, $\forall i = 1,...,K$. However, the fundamental question lies in that: are the above results achievable via the antenna position optimizations? We will perform examinations in the follows.

Note that the conditions $\left| {{\bf{g}}_i^H({\bf{x}}){{\bf{g}}_j}({\bf{x}})} \right|/N = 1/\sqrt {d_{{J_i}}^\tau d_{{J_j}}^\tau } $, $\forall i,j = 1,...,K,i \ne j$, are equivalent to the following equations
\begin{equation}
\setcounter{equation}{75}
\begin{split}{}
\left\{ {\begin{array}{*{20}{c}}
{\frac{{\left| {\sum\nolimits_{i = 1}^N {{e^{j\frac{{2\pi }}{\lambda }{x_i}(\sin {\varphi _1} - \sin {\varphi _2})}}} } \right|}}{N} = 1}\\
{\frac{{\left| {\sum\nolimits_{i = 1}^N {{e^{j\frac{{2\pi }}{\lambda }{x_i}(\sin {\varphi _1} - \sin {\varphi _3})}}} } \right|}}{N} = 1}\\
 \vdots \\
{\frac{{\left| {\sum\nolimits_{i = 1}^N {{e^{j\frac{{2\pi }}{\lambda }{x_i}(\sin {\varphi _{K - 1}} - \sin {\varphi _K})}}} } \right|}}{N} = 1}
\end{array}} \right.,
\end{split}
\end{equation}
which can be further equivalently transformed as
\begin{equation}
\begin{split}{}
\left\{ {\begin{array}{*{20}{c}}
{{e^{j\frac{{2\pi }}{\lambda }{x_1}(\sin {\varphi _i} - \sin {\varphi _j})}} = 1,\forall i,j = 1,...,K,i \ne j}\\
{{e^{j\frac{{2\pi }}{\lambda }{x_2}(\sin {\varphi _i} - \sin {\varphi _j})}} = 1,\forall i,j = 1,...,K,i \ne j}\\
 \vdots \\
{{e^{j\frac{{2\pi }}{\lambda }{x_N}(\sin {\varphi _i} - \sin {\varphi _j})}} = 1,\forall i,j = 1,...,K,i \ne j}.
\end{array}} \right.
\end{split}
\end{equation}

Up to now, we can determine that the ideal antenna positions (denoted as ${{\bf{x}}^*}$) that make all equations in (76) establish can be expressed as
\begin{itemize}
\item[$\bullet$] $x_1^* = 0$.

\item[$\bullet$] $x_2^*$ is the least common multiple of the terms $\frac{\lambda }{{\left| {\sin {\varphi _i} - \sin {\varphi _j}} \right|}},\forall i,j = 1,...,K,i \ne j$.

\item[$\bullet$] ...

\item[$\bullet$] $x_N^*$ is the least common multiple of the terms $\frac{{(N - 1)\lambda }}{{\left| {\sin {\varphi _i} - \sin {\varphi _j}} \right|}},\forall i,j = 1,...,K,i \ne j$.
\end{itemize}

Note that for the convenience of calculating the above least common multiples, it is more advisable to approximate $\frac{\lambda }{{\left| {\sin {\varphi _i} - \sin {\varphi _j}} \right|}}$, $\forall i,j = 1,...,K,i \ne j$, such as only retain one decimal place for these terms.

Via the above analysis, we determine that there exist ideal antenna positions that can make the jamming effects globally optimal. Based on this conclusion, the SINR of ${\rm{R}}_i$ in (4) can be directly simplified as
\begin{equation}
\begin{split}{}
{\gamma _i}({P_i},{\bf{w}}_J^{{\rm{MRT}}},{{\bf{x}}^*}) = \frac{{{P_i}{{\left| {{\bf{w}}_{i,{\rm{ZF}}}^H{{\bf{h}}_i}} \right|}^2}}}{{{Q_J}N/d_{{J_i}}^\tau  + {\sigma ^2}}}.
\end{split}
\end{equation}

Based on (77), when the MAJ aims to reduce the sum rate of the suspicious communications, the performance lower bound would be
\begin{equation}
\begin{split}{}
{\rm{L}}{{\rm{B}}^{\rm{sumrate}}} = \sum\nolimits_{i = 1}^K {{{\log }_2}\left( {1 + {\gamma _i}(P_i^*,{\bf{w}}_J^{{\rm{MRT}}},{{\bf{x}}^*})} \right)},
\end{split}
\end{equation}
\noindent where $P_i^* = \max \left( {\eta  - \frac{{{Q_J}N/d_{{J_i}}^\tau  + {\sigma ^2}}}{{{{\left| {{\bf{w}}_{i,{\rm{ZF}}}^H{{\bf{h}}_i}} \right|}^2}}},0} \right)$ and $\eta $ is the water-filling level which satisfies $\sum\nolimits_{i = 1}^K {P_i^*}  = {P_{{\rm{sum}}}}$.

On the other hand, when the MAJ aims to reduce the fairness of the suspicious communications, the performance bound based on (39) can be determined as
\begin{equation}
\begin{split}{}
{\rm{L}}{{\rm{B}}^{{\rm{minrate}}}} = {\log _2}\left( {1 + \frac{{{P_{{\rm{sum}}}}}}{{\sum\nolimits_{i = 1}^K {\frac{{{Q_J}N/d_{{J_i}}^\tau  + {\sigma ^2}}}{{{{\left| {{\bf{w}}_{i,{\rm{ZF}}}^H{{\bf{h}}_i}} \right|}^2}}}} }}} \right).
\end{split}
\end{equation}

\section{Simulation Results}
In this section, numerical results are presented to evaluate the effectiveness of our proposed schemes.
The simulation setups are illustrated in Fig. 2, where we consider a 2-D topology: the ST is located at $(0,0)$ m, the $i$-th SR, i.e, ${\rm{R}}_i$, is located at $(r,r\tan {\theta _i})$ m, with ${\theta _i} \in \left[ { - \pi /2,\pi /2} \right]$ denoting the corresponding AoA. The MAJ is located at $(r + {r_d},0)$ m. As a result,
 \begin{itemize}
\item[$\bullet$]  The distance between the ST and ${\rm{R}}_i$ is ${d_{{S_i}}} = \frac{r}{{\cos {\theta _i}}}$.

  \item[$\bullet$] The distance between the MAJ and ${\rm{R}}_i$ is ${d_{{J_i}}} = \sqrt {r_d^2 + {{(r\tan {\theta _i})}^2}} $, and the corresponding AoA is ${\varphi _i} = \arctan \left( {\frac{{r\tan {\theta _i}}}{{{r_d}}}} \right)$.
\end{itemize}

 Unless otherwise stated, the system parameters are set as
  \begin{itemize}
\item[$\bullet$]   $r = r_d = 10^3$ m.

  \item[$\bullet$] The number of antennas at the ST is $M = 5$.

  \item[$\bullet$] The number of SRs is $K = 4$.

  \item[$\bullet$] The number of antennas at the MAJ is $N = 5$.

  \item[$\bullet$] The path loss exponent is $\tau  = 3$.

  \item[$\bullet$] The noise power is ${\sigma ^2} = {10^{ - 9}}$ dBm.

    \item[$\bullet$] The signal wavelength $\lambda $ is normalized as one without loss of generality.

    \item[$\bullet$] The minimum spacing between any two adjacent MAs is ${d_{\min }} = \lambda /2 = 1/2$.

     \item[$\bullet$] The positive parameter for the log-sum-exp function is set as $\chi  = 5$.
\end{itemize}

Furthermore, all other relevant parameters will be clearly specified in each figure. In the next, we will comprehensively investigate the achievable performance of our proposed schemes corresponding to the above two scenarios.

\subsection{Convergence Analysis of Proposed Algorithms 1 and 2}
First, we present the convergence behavior of our proposed Algorithm 1 in Fig. 3, where we set ${P_{\rm{{sum}}}} = {Q_J} = 10$ dBm. From Fig. 3, we can observe that the suspicious sum rate decreases w.r.t. the inner/outer iteration index and the curves arrive at the stationary value after about dozens of iterations, which thus validate the convergence analysis in Section III.

Second, we present the convergence behavior of our proposed Algorithm 2 in Fig. 4, where we set ${P_{{\rm{sum}}}} = 20$ dBm and ${Q_J} = 10$ dBm. From Fig. 4, we can observe that the objective, i.e., ${{\bf{y}}^H}{{\bf{G}}^H}({\bf{x}}){\bf{G}}({\bf{x}}){\bf{y}}$, converges to a stationary value without exceeding ten inner/outer iterations, indicating that our proposed Algorithm 2 is computationally efficient.

\subsection{Performance Comparison}
To perform entire comparisons and verify the effectiveness of our proposed schemes, we further consider five schemes as benchmarks for minimizing the sum rate and the fairness, respectively.
\subsubsection{Benchmarks for Minimizing the Sum Rate}
\begin{itemize}
\item[$\bullet$] \textbf{Random Antenna Positions (RAP)}: Randomly generate antenna positions satisfying the constraints in (6c) and (6d) for $10^2$ independent realizations. The MAJ optimizes the jamming beamforming by solving (P2.1) given each random antenna positions, and then selects the minimum output.

  \item[$\bullet$] \textbf{FPA Array}: There are $N$ antennas at the MAJ, and the distance between arbitrary two adjacent antennas is fixed as $\lambda /2$. Then, the MAJ just optimizes its jamming beamforming by solving (P2.1).

 \item[$\bullet$] \textbf{Antenna Selection (AS)}: The MAJ is equipped with FPA-based uniform linear array with $N + 3$ antennas. Then, it randomly selects $N$ antennas with ${\rm{C}}_{N + 3}^N$ realizations. Given each realization, the MAJ optimizes the jamming beamforming by solving (P2.1) and then selects the minimum output.

 \item[$\bullet$] \textbf{Fixed Beamforming with Exhaustive Antenna Positions (FB+EAP)}: Exhaustively generate antenna positions satisfying the constraints in (6c) and (6d) for $10^4$ independent realizations. Given each realization ${\bf{x}}$, the MAJ fixes its jamming beamforming as ${{\bf{w}}_J}({\bf{x}}) = \sum\nolimits_{i = 1}^K {{{\bf{g}}_i}({\bf{x}})} /\left\| {\sum\nolimits_{i = 1}^K {{{\bf{g}}_i}({\bf{x}})} } \right\|$. The MAJ computes the objective of (P2) given each ${\bf{x}}$ and ${{\bf{w}}_J}({\bf{x}})$ and then selects the minimum output.

\item[$\bullet$] \textbf{Rotatable uniform linear array (RULA) \cite{RULA}}:  The MAJ is equipped with RULAs with $N$ antennas, spaced by $\lambda /2$. The rotation of the RULA is quantized into $50$ discrete angles in the range of $[ - \pi /5,\pi /5]$. Given each rotated angle, the MAJ optimizes its jamming beamforming by solving (P2.1) and then selects the minimum output.
\end{itemize}

\subsubsection{Benchmarks for Minimizing the Fairness}
\begin{itemize}
\item[$\bullet$] \textbf{RAP}: Randomly generate antenna positions satisfying the constraints in (6c) and (6d) for $10^2$ independent realizations. Then, substitute each realization into (P3.1), compute the objective value and then select the maximum output.

  \item[$\bullet$] \textbf{FPA Array}: There are $N$ antennas at the MAJ, and the distance between arbitrary two adjacent antennas is fixed as $\lambda /2$. As antenna positions are fixed, this scheme produces a fixed principal eigenvalue of ${{\bf{G}}^H}({\bf{x}}){\bf{G}}({\bf{x}})$. Therefore, the minimum suspicious rate can be immediately obtained.

 \item[$\bullet$] \textbf{AS}: The MAJ is equipped with FPA-based uniform linear array with $N + 3$ antennas. Then, it randomly selects the ``best'' $N$ antennas with ${\rm{C}}_{N + 3}^N$ realizations in order to maximize the objective of (P3.1).

 \item[$\bullet$] \textbf{FB+EAP}: Exhaustively generate antenna positions satisfying the constraints in (6c) and (6d) for $10^4$ independent realizations. Given each realization ${\bf{x}}$, the MAJ fixes its jamming beamforming as ${{\bf{w}}_J}({\bf{x}}) = \sum\nolimits_{i = 1}^K {{{\bf{g}}_i}({\bf{x}})} /\left\| {\sum\nolimits_{i = 1}^K {{{\bf{g}}_i}({\bf{x}})} } \right\|$. The MAJ computes the objective of (P3) (instead of (P3.1)) given each ${\bf{x}}$ and ${{\bf{w}}_J}({\bf{x}})$ and then selects the maximum output.

     \item[$\bullet$] \textbf{RULA}: Similarly, the rotation of the RULA is quantized into $50$ discrete angles in the range of $[ - \pi /5,\pi /5]$. Substitute each angle realization into (P3.1), compute the objective value and then select the maximum output.
\end{itemize}

\begin{table*}[t]
\caption{\textcolor{blue}{Complexity Comparisons of Different Schemes}}
\renewcommand\arraystretch{1.35}
\normalsize
\centering
\begin{tabular}[l]{|c|c|c|c|c|c|c|}
 \hline
 &Minimizing the Sum Rate&Minimizing the Fairness\\
 \hline
  Proposed &${\cal O}\left( {{I_{{\rm{iter}},{\rm{out}}}}\left( {{I_{{\rm{iter}},1}}{{(K + 1 + 2N)}^{3.5}} + {I_{{\rm{iter}},2}}{{(K + 1 + N)}^{3.5}}} \right)} \right)$& ${\cal O}({I_{{\rm{iter}},{\rm{out}}}}\left( {{I_{{\rm{iter}},1}} + {I_{{\rm{iter}},2}}{N^{3.5}}} \right))$\\
 \hline
RAP&${\cal O}\left( {{{10}^2} \times {I_{{\rm{iter}},1}}{{(K + 1 + 2N)}^{3.5}}} \right)$&${\cal O}\left( {{{10}^2} \times {I_{{\rm{iter}},{\rm{prineig}}}}{N^2}} \right)$\\
 \hline
FPA Array&${\cal O}\left( {{I_{{\rm{iter}},1}}{{(K + 1 + 2N)}^{3.5}}} \right)$& ${\cal O}\left( {{I_{{\rm{iter}},{\rm{prineig}}}}{N^2}} \right)$\\
\hline
AS&${\cal O}\left( {{\rm{C}}_{N + 3}^N{I_{{\rm{iter}},1}}{{(K + 1 + 2N)}^{3.5}}} \right)$& ${\cal O}\left( {{\rm{C}}_{N + 3}^N{I_{{\rm{iter}},{\rm{prineig}}}}{N^2}} \right)$\\
 \hline
FB+EAP&${\cal O}\left( {{{10}^4} \times K\log (K)} \right)$& ${\cal O}\left( {{{10}^4}} \right)$\\
 \hline
RULA&${\cal O}\left( {50 \times {I_{{\rm{iter}},1}}{{(K + 1 + 2N)}^{3.5}}} \right)$& ${\cal O}\left( {50 \times {I_{{\rm{iter}},{\rm{prineig}}}}{N^2}} \right)$\\
 \hline
 \end{tabular}
{
 \label{tab:tb1}
}
\end{table*}

\begin{figure}
\centering
\includegraphics[width=7.9cm]{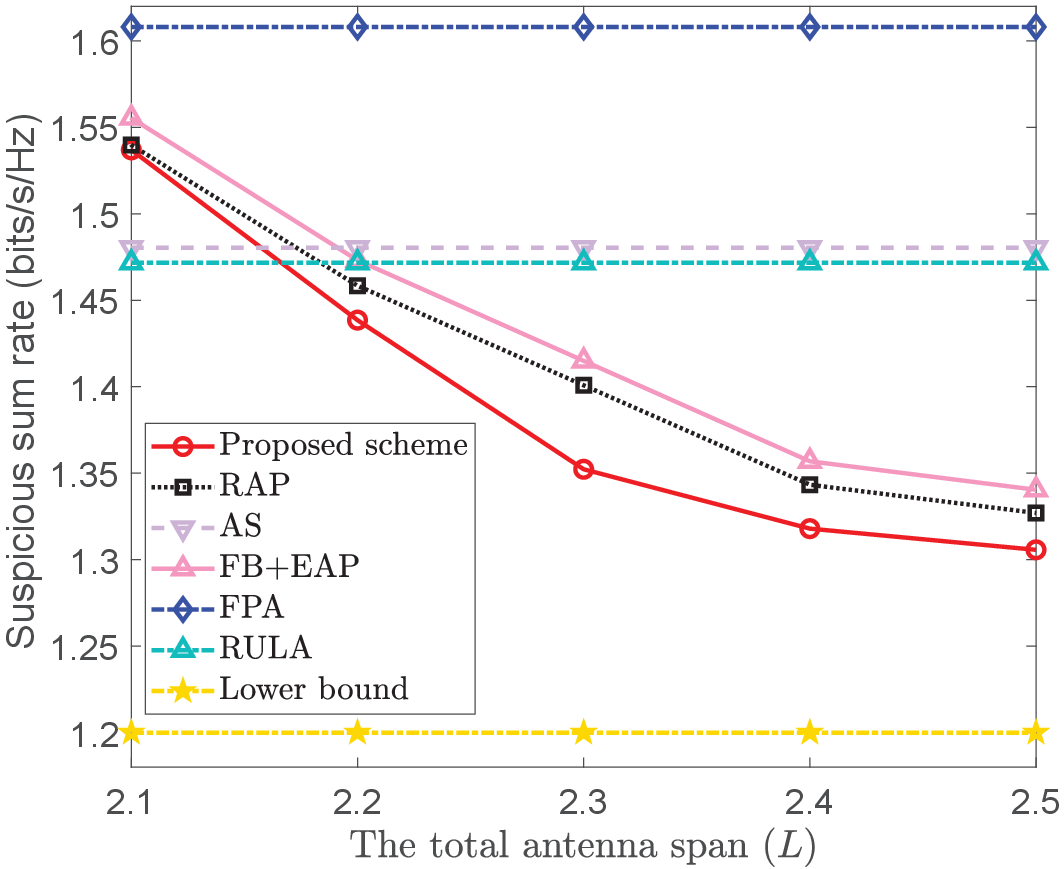}
\captionsetup{font=small}
\caption{Suspicious sum rate w.r.t. the total antenna span ($L$) at the MAJ, where ${P_{{\rm{sum}}}} = {Q_J} = 10$ dBm.} \label{fig:Fig1}
\end{figure}

\begin{figure}
\centering
\includegraphics[width=7.9cm]{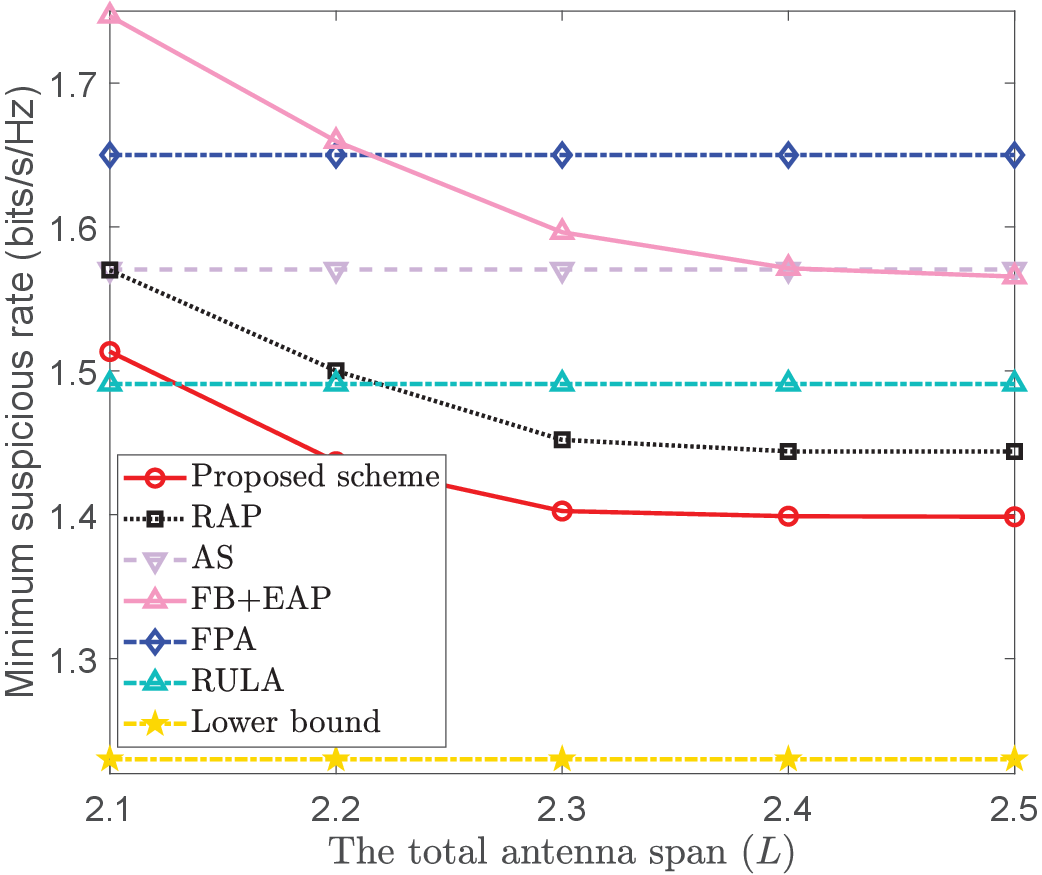}
\captionsetup{font=small}
\caption{Minimum suspicious rate w.r.t. the total antenna span ($L$) at the MAJ, where ${P_{{\rm{sum}}}} = 20$ dBm and ${Q_J} = 10$ dBm.} \label{fig:Fig1}
\end{figure}

Table I presents the complexity comparisons between our proposed scheme and the other five benchmark schemes. For the RAP, FPA Array, AS and RULA schemes, the term ${{I_{{\rm{iter}},{\rm{prineig}}}}{N^2}}$ represents the complexity of computing the principal eigenvalue of ${\bf{G}}({\bf{x}}){{\bf{G}}^H}({\bf{x}})$ with the given ${\bf{x}}$ or rotated angle. Here, ${{I_{{\rm{iter}},{\rm{prineig}}}}}$ denotes the number of iterations required by eigenvalue computation methods, such as the power iteration or Lanczos algorithm. In addition, the complexity of the FB+EAP scheme for sum-rate minimization is derived as follows: for each realization ${\bf{x}}$ and ${{\bf{w}}_J}({\bf{x}})$, the MAJ must determine the optimal water-filling level $\eta $ of the suspicious system. This step incurs a computational complexity of ${K\log (K)}$.

The performance of our proposed schemes and the above benchmarks w.r.t. the total antenna span ($L$) at the MAJ is presented in Fig. 5 and Fig. 6, from which we observe that: \textbf{i)} As $L$ increases, antennas at the MAJ have the larger movement spaces to reconfigure jamming channels more flexibly for better adapting to the jamming beamforming. Based on this fact, the suspicious sum rate and the minimum suspicious rate caused by our proposed schemes monotonically decrease w.r.t. $L$. In addition, the suspicious sum rate and the minimum suspicious rate converge to a steady state as $L$ becomes enough large. This is mainly due to that the arbitrary element in the jamming channel vector is closely related to the cosine and sine of the antenna position, and the cosine and sine functions have periodic characteristics. This observation implies that it is not necessary to expand $L$ infinitely and only a finite $L$ is enough to achieve the satisfactory performance; \textbf{ii)} For the scheme of RAP, although the jamming beamforming is carefully designed, the ``best'' antenna positions are determined from multiple random realizations without specific optimization rules. Therefore, the suspicious sum rate and the minimum suspicious rate caused by RAP are relatively higher than our proposed schemes; \textbf{iii)} For the scheme of AS, the ``best'' $N$ antennas can be selected from $N + 3$ discrete locations. The movement freedom of antenna positions using AS is relatively higher than our proposed schemes when $L$ is small. However, when $L$ increases, the situation will become opposite. This is due to that using our proposed schemes, antennas at the MAJ can move in a continuous manner within a larger space; \textbf{iv)} Using FB+EAP, although antenna positions at the MAJ are optimized using an exhaustive manner, the jamming beamforming is just simply setting without carefully optimization. Therefore, the corresponding performance is also unfavorable compared to our proposed schemes; \textbf{v)} Based on RULA, the antenna array at the MAJ can only be reconfigured by rotation, which implies that the spatial channel variation cannot be fully exploited. Therefore, the scheme of RULA is moderate compared to our proposed one; \textbf{vi)} For the scheme of FPA, since antenna positions at the MAJ are fixed, i.e., there is no any additional freedom that can be exploited, such scheme achieves the worst performance; \textbf{vii)} When $L$ is large, e.g. 2.5 in Fig. 5 and Fig. 6, a certain gap still exists between the performance achieved by the proposed scheme and its lower bound. This is primarily because the ideal antenna span required to achieve the performance lower bound is, in fact, very large.

 \begin{figure}
\centering
\includegraphics[width=7.9cm]{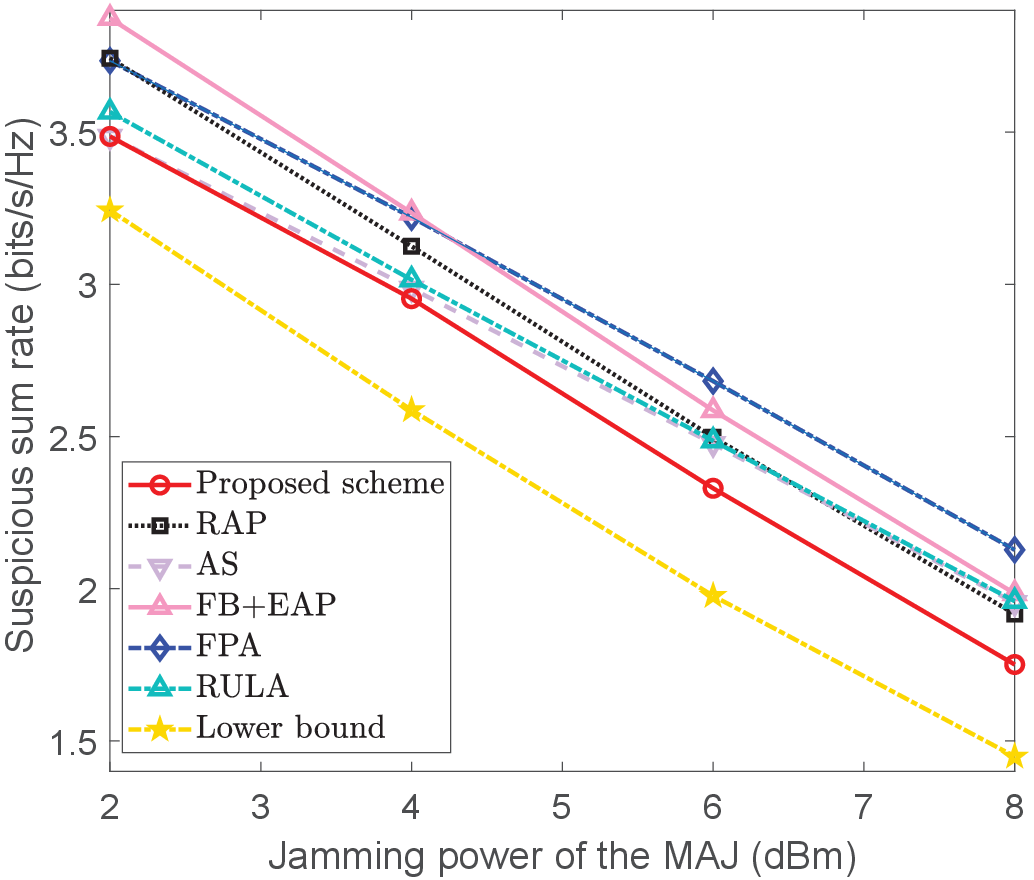}
\captionsetup{font=small}
\caption{Suspicious sum rate w.r.t. the jamming power ($Q_J$) of the MAJ, where ${P_{{\rm{sum}}}} = 10$ dBm and $L = 2.3$.} \label{fig:Fig1}
\vspace{-10pt}
\end{figure}

 \begin{figure}
\centering
\includegraphics[width=7.9cm]{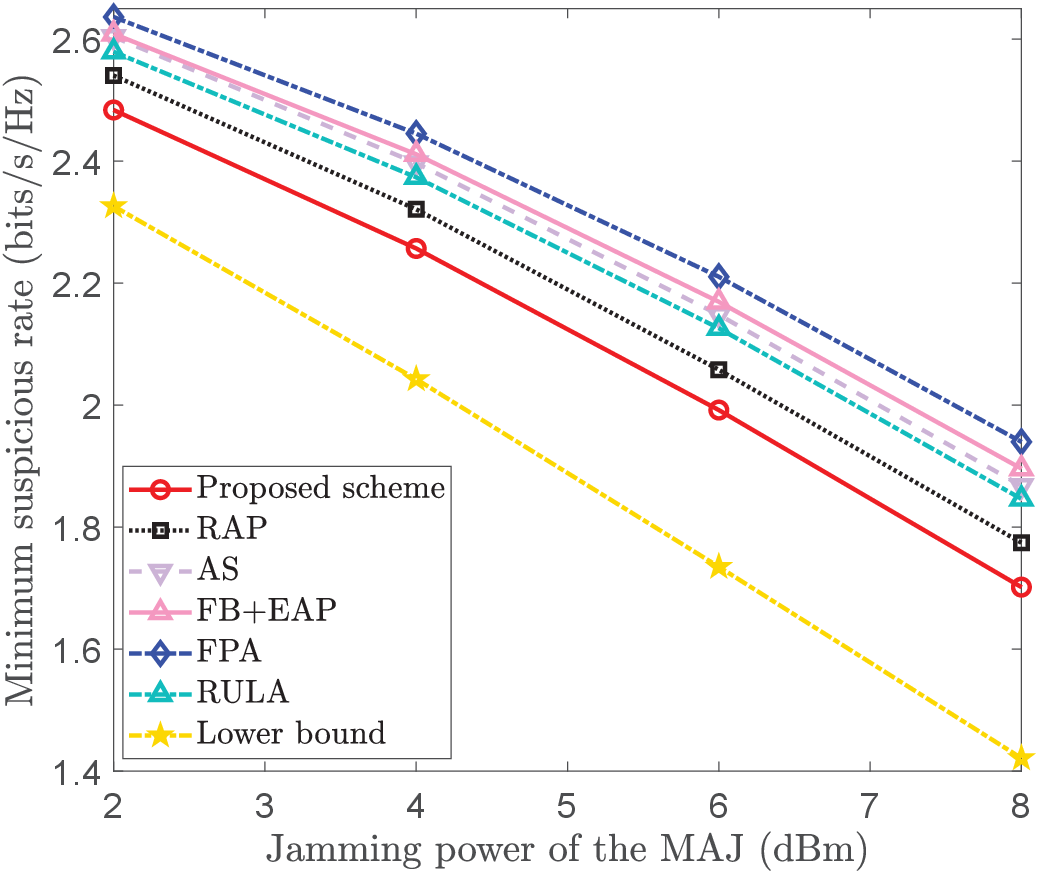}
\captionsetup{font=small}
\caption{Minimum suspicious rate w.r.t. the jamming power of the MAJ, where ${P_{{\rm{sum}}}} = 20$ dBm and $L = 2.3$.} \label{fig:Fig1}
\vspace{-10pt}
\end{figure}

 \begin{figure}
\centering
\includegraphics[width=7.9cm]{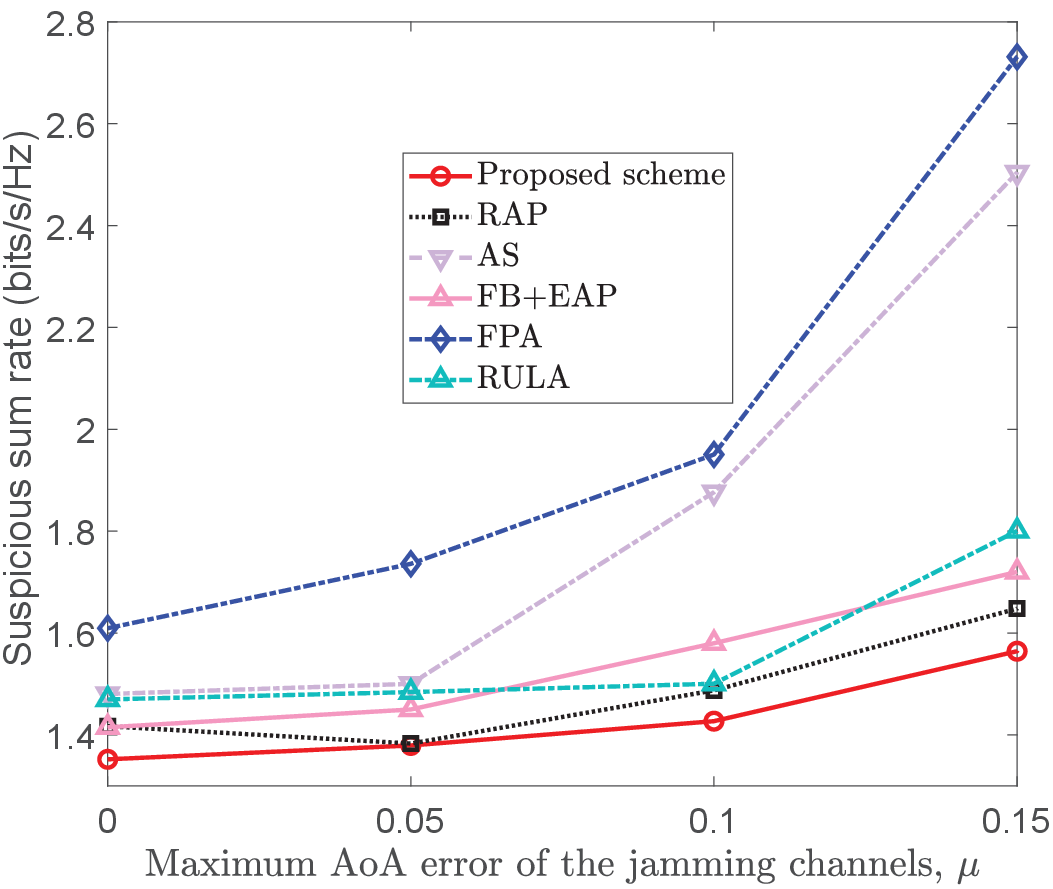}
\captionsetup{font=small}
\caption{Suspicious sum rate w.r.t. the maximum AoA error, where ${P_{{\rm{sum}}}} = Q_J = 10$ dBm and $L = 2.3$.} \label{fig:Fig1}
\vspace{-10pt}
\end{figure}

 \begin{figure}
\centering
\includegraphics[width=7.9cm]{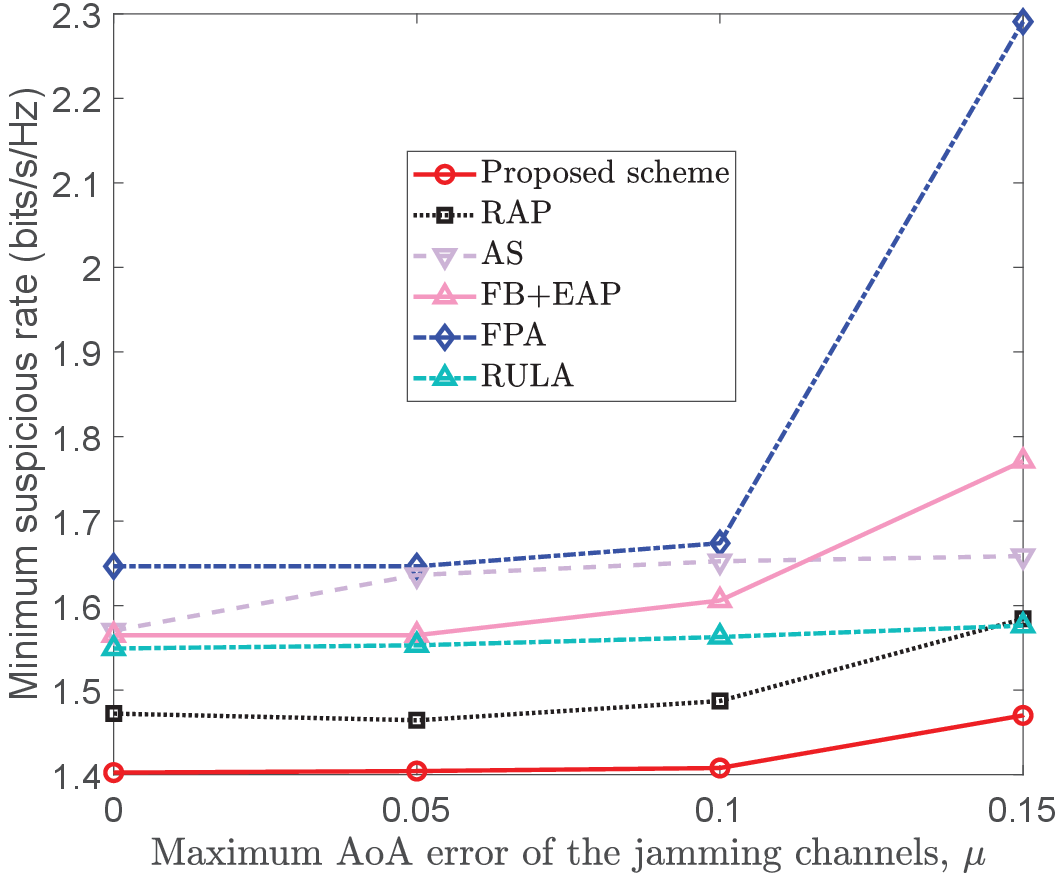}
\captionsetup{font=small}
\caption{Minimum suspicious rate w.r.t. the maximum AoA error, where ${P_{{\rm{sum}}}} = 20$ dBm, ${Q_J} = 10$ dBm and $L = 2.3$.} \label{fig:Fig1}
\vspace{-10pt}
\end{figure}

Fig. 7 and Fig. 8 respectively illustrate the suspicious sum rate and minimum suspicious rate of six schemes w.r.t. the jamming power of the MAJ, from which we can observe that: As $Q_J$ increases, the jamming effect to each suspicious receiver becomes stronger, no matter how the ST optimizes its power allocations. Therefore, obviously the suspicious sum rate and the minimum suspicious rate become lower w.r.t. $Q_J$.

%
%
%
%

To show the robustness of our proposed schemes, we further present the suspicious sum rate and the minimum suspicious rate w.r.t. the maximum AoA error as shown in Fig. 9 and Fig. 10, respectively. Here, the maximum AoA error $\mu $ means that for the MAJ, it cannot estimate the exact ${\theta _i}$ and ${\varphi _i}$, $i = 1,...,K$, and just knows that such AoAs follow uniform distributions, i.e., ${\theta _i} \sim {\cal U}[{\widehat \theta _i} - \mu /2,{\widehat \theta _i} + \mu /2]$ and ${\varphi _i} \sim {\cal U}[{\widehat \varphi _i} - \mu /2,{\widehat \varphi _i} + \mu /2]$, with ${\widehat \theta _i}$ and ${\widehat \varphi _i}$ denoting the estimated AoAs, $i = 1,...,K$. The results of Fig. 9 and Fig. 10 are obtained based on the following setups: \textbf{i)} The jamming beamforming and antenna positions at the MAJ are optimized based on the estimated AoAs, i.e., ${{{\widehat \theta }_i}}$ and ${{{\widehat \varphi }_i}}$, $i = 1,...,K$; \textbf{ii)} Then, the resulted suspicious sum rate and the minimum suspicious rate are calculated by substituting the optimized variables and the actual AoAs into the objective. From Fig. 9 and Fig. 10 we can clearly observe that: \textbf{i)} The resulted suspicious sum rate and the minimum suspicious rate by our proposed schemes, RAP and FB+EAP are slightly increasing w.r.t. the maximum AOA error. For instance, using our proposed schemes, compared with the situation where $\mu  = 0$, when $\mu  = 0.1$, the increment of the suspicious sum rate or the minimum suspicious rate does not exceed $5\% $; \textbf{ii}) While for the scheme of FPA, as the maximum AoA error increases, the suspicious sum rate and the minimum suspicious rate increases rapidly. Therefore, it can be concluded that our proposed schemes are robust to the maximum AoA error.

\section{Conclusion}
This paper exploited the MA-enabled jammer to subvert flexible multiuser suspicious communications. The MAJ aimed to jointly optimize its jamming beamforming and antenna positions to minimize the benefits (sum rate or minimum rate) of the suspicious communications, while the ST can adaptively optimize its power allocations to moderate unfavorable interference caused by the MAJ, which caused certain confusions to the MAJ about its exact jamming effects. Facing this challenge, the optimal behaviors of the ST given the MAJ's actions were first inferred, based on which we arrived at two simplified problems which were then solved using AO algorithms. Also, insightful conclusions about the deployment rule of antenna positions at the MAJ were derived by focusing on the special case of two SRs. Ideal antenna position deployments which can achieve the globally performance bounds were also analyzed theoretically. Numerical results demonstrated the effectiveness of our proposed schemes compared to competitive benchmarks.

Based on the findings of this work, several promising directions warrant further investigation:

Robust Optimization under Imperfect CSI: The performance of the proposed scheme likely relies on the quality of CSI. A critical future direction is to design robust jamming algorithms that account for imperfect, partial, or delayed CSI. This would ensure the jamming effectiveness remains high even in the face of significant channel uncertainty, making the system more practical for real-world deployment.

Energy Efficiency and Hardware Constraints: While this work focused on performance bounds, future studies could incorporate practical constraints such as the energy consumption of antenna movement and the physical limitations of the MA platform. Investigating the trade-off between jamming performance and energy efficiency, and designing corresponding green jamming strategies, would be of great practical value for sustained operations.

\normalem
\bibliographystyle{IEEEtran}
\bibliography{IEEEabrv,mybib}

\end{document}